# Anisotropic Thermal Transport in
# Tunable Self-Assembled Nanocrystal Supercrystals

Matias Feldman,[1] Charles Vernier,[2] Rahul Nag,[3] Juan J. Barrios-Capuchino,[4] Sébastien Royer,[1] Hervé Cruguel,[1] Emmanuelle Lacaze,[1] Emmanuel Lhuillier,[1] Danièle Fournier,[1] Florian Schulz,[4] Cyrille Hamon,[3] Hervé Portalès,[2] James K. Utterback[1]*

[1] Sorbonne Université, CNRS, Institut des NanoSciences de Paris, 75005 Paris, France
[2] Sorbonne Université, CNRS, MONARIS, 75005 Paris, France
[3] Laboratoire de Physique des Solides, CNRS and Université Paris-Saclay, 91400 Orsay, France
[4] Institute for Nanostructure and Solid State Physics, University of Hamburg, Luruper Chaussee 149, 22761 Hamburg, Germany

*Correspondence to: james.utterback@sorbonne-universite.fr

**Abstract**

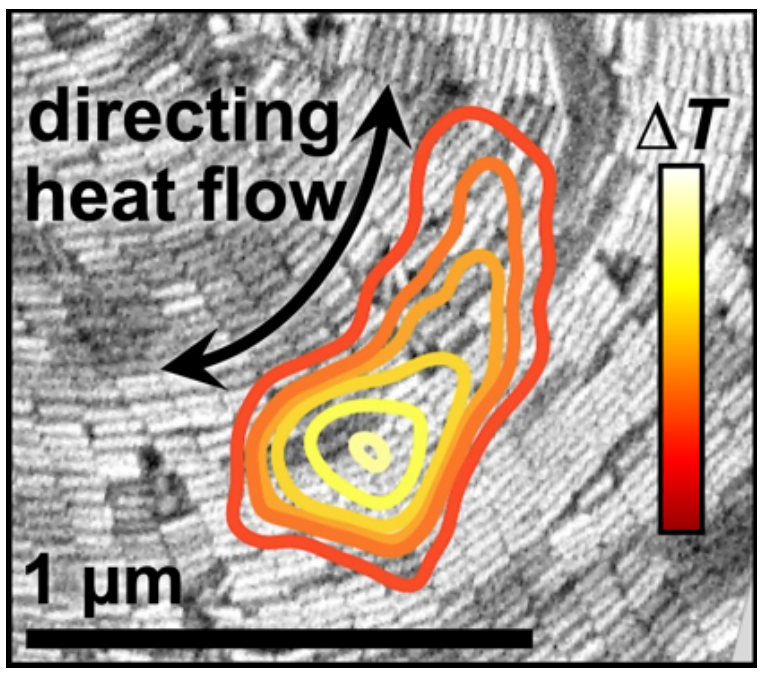

Realizing tunable functional materials with built-in nanoscale heat flow directionality represents a significant challenge that could advance thermal management strategies. Here we use spatiotemporally-resolved thermoreflectance to visualize lateral thermal transport anisotropy in self-assembled supercrystals of anisotropic Au nanocrystals. Correlative electron and thermoreflectance microscopy reveal that nano- to meso-scale heat predominantly flows along the long-axis of the anisotropic nanocrystals, and does so across grain boundaries and curved assemblies while voids disrupt heat flow. We finely control the anisotropy via the aspect ratio of constituent nanorods, and it exceeds the aspect ratio for nano-bipyramid supercrystals and certain nanorod arrangements. Finite element simulations and effective medium modeling rationalize the emergent anisotropic behavior in terms of a simple series resistance model, further providing a framework for estimating thermal anisotropy as a function of material and structural parameters. Self-assembly of colloidal nanocrystals promises an interesting route to direct heat flow in a wide range of applications that utilize this important class of materials.

Supercrystals (SCs) self-assembled from colloidal nanocrystals (NCs) offer a versatile platform for tuning emergent material properties by controlling the constituent building blocks and interfaces. Efforts over the past four decades have enabled the rational assembly of a vast library of ordered NC structures with exceptional control over nanocrystal shape, size, and material composition as well as surface ligand-mediated inter-NC interaction.[1–4] The collective properties of NC SCs enable precisely tunable optical, electronic, phononic, mechanical, and catalytic functionalities relevant to a wide range of applications not possible in bulk materials.[5–9] Controlling nanoscale thermal transport is fundamental to all such applications, as they either inherently generate heat as a byproduct or deliberately harness it for operation. Seminal work by Malen and co-workers and subsequent studies have established that thermal transport in NC solids is dominated by the ligand thermal conductivity, NC–ligand interfacial resistance, and NC volume fraction while being only weakly dependent on NC core material.[10–18] NC ordering also plays an important role, ranging from optimized thermal conductivity of ordered NC SCs[12] to slow—even subdiffusive—transport in the case of strong disorder and voids.[19] This ability of the NC–ligand interface, NC size, and arrangement to modulate thermal transport raises the question: can NC shape and anisotropic assembly induce and tune anisotropic transport on macroscopic scales? Such behavior could present a route to directing heat flow in self-assembled solids.

Accessing the intrinsic thermal transport properties of anisotropic NC SCs is a challenge that simultaneously requires large ordered domains with defined orientations and a local probe of heat flow with direction-specific sensitivity. Most previous reports on the thermal properties of NC solids measured assemblies with varying degrees of disorder,[10,13,14,19–21] and only recent efforts evaluated single SC domains of spherical NCs.[12] These previous studies investigated isotropic thermal transport that is expected for hexagonal packing of spherical NCs or when probing length scales larger than the structural order. In the case of structurally anisotropic NC building blocks such as nanorods and nano-bipyramids, self-assembly protocols achieve highly ordered, oriented SCs with domains extending few to tens of microns.[2,22–24] Optically-based thermal probes such as thermoreflectance[25,26] offer an opportunity to probe such length scales. In particular, spatiotemporally-resolved transient microscopy[19,27–30] is well suited to visualize in-plane transport anisotropy without relying on modeling,[27,31,32] and has recently been applied to image isotropic thermal transport in disordered Au NC films.[14,19] The confluence of these self-assembly and pump-probe microscopy advances presents an opportune moment to investigate the hypothesis of directional thermal transport in anisotropic NC SCs.

Here, we use spatiotemporally-resolved thermoreflectance to reveal the nano- to meso-scale anisotropy of in-plane thermal transport in self-assembled SCs with anisotropic colloidal Au NC building blocks. We optically image heat propagation on a nanosecond timescale with sub-micron resolution in SC domains that extend tens of microns. The expansion appears faster along one axis, and we directly obtain the thermal diffusivity along two orthogonal directions to quantify the thermal transport anisotropy ratio. Correlative electron microscopy and microscopic thermal transport measurements directly confirm that the axis of high thermal conductivity is parallel to the long-axis of the anisotropic nanocrystals. Furthermore, heat flow is guided by the grains of the nanocrystal orientations across tilt grain boundaries and mesoscale curved arrangements, while it is blocked by voids. For nanorod SCs with typical side-to-side centered rectangular packing, the thermal transport anisotropy is tunable, with orthogonal diffusivity ratios increasing monotonically from 1 to 5 with increasing aspect ratio from 1 to 7. In nano-bipyramid SCs and staggered nanorod arrangements that couple transport along the fast and slow axes, the transport anisotropy exceeds the aspect ratio. Packing defects also provide a means to decrease transport along the short axis and thereby increase the anisotropy. By reproducing the experiments with finite element simulations, we show that Fourier's law and geometry are adequate to describe the anisotropic thermal transport in this class of systems. Effective medium approximation calculations adequately describe the data in terms of series resistances and provide a framework to predict such behavior beyond the materials and dimensions of the system studied here. This work establishes self-assembly as a route to achieving single-axis anisotropic thermal transport and even curved thermal routing within a three-dimensional material on nano- to micro-scales with tunable building blocks. This behavior is expected to extend to semiconductor NCs as well and thus offers a potential thermal management strategy in a wide range of NC applications.

# Results

## Visualizing anisotropic thermal transport

We assemble three-dimensional NC SCs with ordered domains ranging from few to tens of microns by solvent evaporation (Figure 1, S1).[2,23,24,33,34] As building blocks, we use colloidal Au NCs with a range of shapes and aspect ratios, including isotropic spheres capped with polystyrene thiol ligands and anisotropic rod- and bipyramid-shaped particles capped with long chain amines (Figure S2, S3, Table S1, S2, see Methods on NC and SC preparation). The Au NCs assemble into three-dimensional SCs with their long-axis lying parallel to the substrate (Figure 1).[23] To locally visualize the lateral thermal transport in these structures, we use spatiotemporally-resolved thermoreflectance microscopy—also known as stroboscopic optical scattering microscopy (stroboSCAT) as developed by Delor et al.[27,28,30] In this experiment, we optically generate a diffraction-limited Gaussian heat pulse and track its evolution in space and time by imaging the resulting temperature-induced change in reflectance ($\Delta R/R$) of a wide-field probe light pulse (Figure 1a, see Methods on stroboSCAT). Because the central $\Delta R/R$ profile is proportional to the transient temperature change of Au,[35–37] we refer to it as the transient temperature profile in what follows.

For SCs with anisotropic NC building blocks, the expansion of the temperature profile over time is asymmetric, expanding faster along one axis (Figure 1b). We attribute this behavior to thermal transport anisotropy in which the thermal diffusivity (or conductivity) is larger along one direction owing to the anisotropy of the constituent NCs of the SC (Figure 1c). To characterize this anisotropy, we fit the temperature profiles to a two-dimensional elliptical Gaussian of width $\sigma_{\parallel}$ along the major axis and $\sigma_{\perp}$ along the minor axis as a function of time $t$, calculate the mean-squared expansion curves, and thereby obtain the axis-specific thermal diffusivities of the composite NC SC medium: $\sigma_{\parallel,\perp}^2(t) - \sigma_{\parallel,\perp}^2(t_0) = 2D_{\parallel,\perp}t$ (Figure 1d). Mean values of $D_{\perp}$ and $D_{\parallel}$ for all measured samples fall in the range of 0.3–1.7 × $10^{-3}$ cm$^2$ s$^{-1}$ and 0.8–9.7 × $10^{-3}$ cm$^2$ s$^{-1}$, respectively (Table S3). We estimated the corresponding thermal conductivities to be 0.07–0.3 W m$^{-1}$ K$^{-1}$ along the minor axis—consistent with the order of magnitude previously measured for related NC solids[10]—and up to 1.7 W m$^{-1}$ K$^{-1}$ along the major axis for the longest nanorod sample (Table S4). Importantly, we quantify the thermal transport anisotropy through the ratio of diffusion coefficients for the major and minor axes, which for the example of Au nanorods with a length-to-diameter aspect ratio of 4:1 gives $D_{\parallel}/D_{\perp} = 3.2 \pm 0.3$ (Figure 1d). We find that this thermal anisotropy persists out to at least 10 μm within a single SC domain despite the potential cumulative effects of defects (Figure S4).

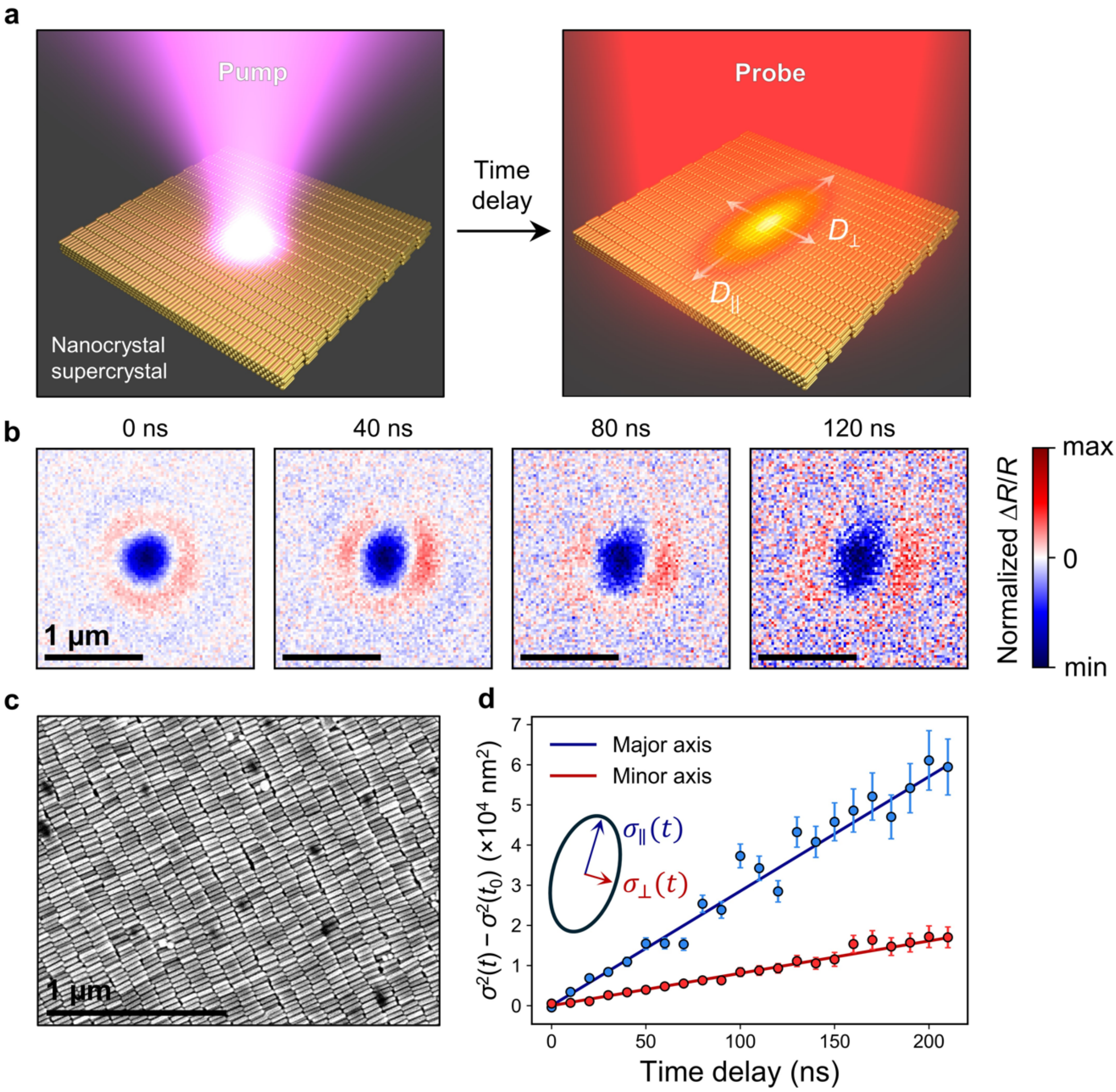

**Figure 1. Spatiotemporal visualization of anisotropic thermal transport in nanocrystal supercrystals.** **a**, Schematic depiction of stroboSCAT experiment wherein the sample is excited by a focused 405 nm pump light pulse (340 nm full-width at half-maximum) then imaged by a reflected wide-field 640 nm probe pulse after some time delay. **b**, Representative transient reflectance image time series in a supercrystal of 4:1 Au nanorods. The profile is initially a symmetric Gaussian shape and becomes elliptical over time due to anisotropic transport. **c**, Representative SEM image of a 4:1 Au nanorod SC. **d**, Mean-squared expansion curves of major and minor axes as a function of time for data in (**b**). Data are fit to $2Dt$ to obtain the ratio $D_\parallel/D_\perp = 3.2 \pm 0.3$. Error bars represent the fitting uncertainty of $\sigma_{\parallel,\perp}^2$ for the two-dimensional Gaussian fits to the profiles exemplified in (**b**).

## Correlative transport and electron microscopy measurements

To directly reveal the influence of NC packing and mesoscale structure on anisotropic heat flow, we perform correlative scanning electron microscopy (SEM) and stroboSCAT on a SC composed of nanorods with a 4:1 aspect ratio (Figure 2a,b, Figure S5). Figure 2c-g show how heat propagates over time for various NC packing motifs and superimposes the late-time temperature profiles from stroboSCAT over SEM images collected in the same locations. In the case of well-ordered nanorod packing, the temperature profile expands most rapidly in the direction parallel to the long-axis of the nanorods (Figure 2c). This confirms that the effective thermal conductivity of the SC is higher along the nanorod long-axis than it is along the short-axis. In addition to the local anisotropy imposed by the NC building blocks, we visualize mesoscale modulation due to the SC structure. Exciting at a tilt grain boundary leads to wings that extend parallel to the nanorod orientations on either side (Figure 2d). Similarly, in liquid-crystalline-like regions where there is progressive curvature in the nanorod orientations,[22,23] heat flow results in a crescent-shaped profile at long times (Figure 2e). This suggests a physical picture in which nanoscale heat transport is modulated on the scale of individual NCs, following the trajectory guided by the local nanorod orientations.

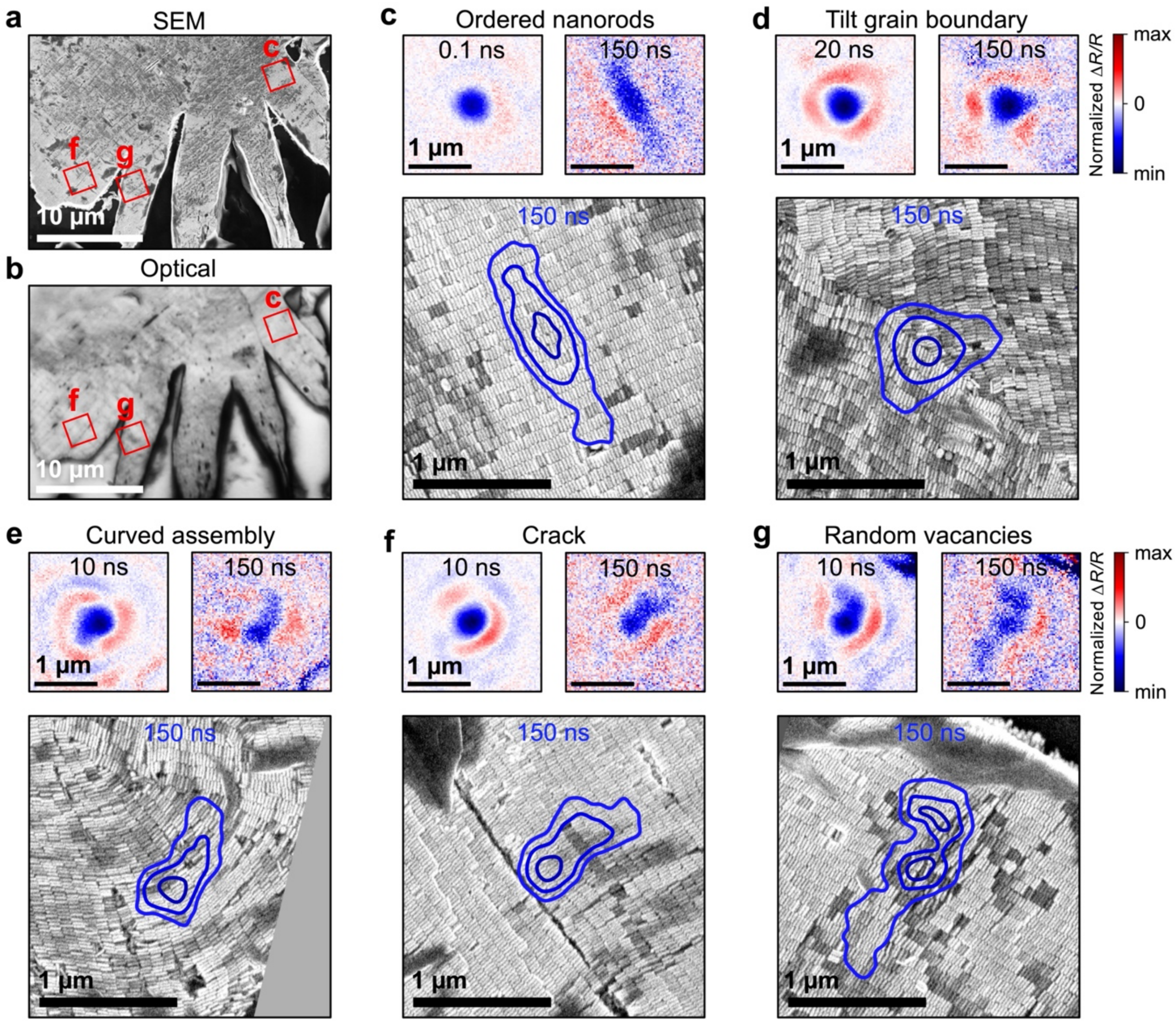

**Figure 2. Correlation between nanocrystal packing arrangement and thermal transport.** Representative SEM, **a**, and optical, **b**, images of the same region of a 4.0:1 Au–CTAB nanorod supercrystal sample. Examples of locations where spatiotemporally-resolved thermoreflectance measurements were performed are indicated with red squares corresponding to the data sets shown in panels (**c**), (**g**) and (**f**). **c-g**, Normalized-amplitude stroboSCAT images at an early and late time before and after heat propagation and correlative late-time $\Delta R/R$ contour overlaying SEM image for supercrystal regions with (**c**), ordered, close-packed nanorods, (**d**), a tilt grain boundary, (**e**), liquid-crystalline-like curved packing, (**f**), a large crack in the supercrystal, and (**g**), a high density of randomly positioned nanorod void defects. Corresponding optical microscopy images appear in Figure S5.

We further use this powerful approach of correlative SEM and stroboSCAT to access structure–function relationships of defects and nanoscale heat transport. Exciting next to a crack in the film shows that heat does not flow across such morphological features, expanding primarily in one direction away from it (Figure 2f). Similarly, in regions with a high density of randomly distributed vacancies, the temperature profile develops an irregular shape as it navigates the region (Figure 2g). These observations suggest that heat conduction primarily goes around voids,[19] relying on direct contact between neighboring NCs and their ligands on the timescale of the measurement.

## Controlling thermal anisotropy through aspect ratio

We explore the ability to control the thermal transport anisotropy via the NC aspect ratio (Figure 3a). In a series of SC samples self-assembled from nanospheres of aspect ratio 1 (Figure 3b), nanorods with aspect ratios ranging from 2 to 7 (Figure 3c,e, S1, Table S2), as well as nano-bipyramids of aspect ratio 3.2 (Figure 3d), we measure $D_{||}$ and $D_{\perp}$ in multiple locations (Table S3, S4, Figure S6). Despite fluctuations in the magnitudes of $D_{||}$ and $D_{\perp}$ that are likely due to variations in ligand density and SC packing density between different NC batches and self-assembly preparations (Table S3), the trend in thermal diffusivity ratio is robust. We find that the thermal transport anisotropy, $D_{||}/D_{\perp}$, is 1 for nanospheres, then it increases monotonically with nanorod aspect ratio up to a value of 5 (Figure 3a). The slope of this relationship is less than one for the typical side-to-side nanorod packing structure (Figure 3c), yet is significant enough to result in considerable anisotropy. The nano-bipyramids exhibit an anisotropy exceeding the aspect ratio. Taken together, these results demonstrate that the degree of anisotropy can be tuned with a high degree of predictability, speaking to the robustness of this phenomenon in this class of systems. Below we develop a model for the above trends.

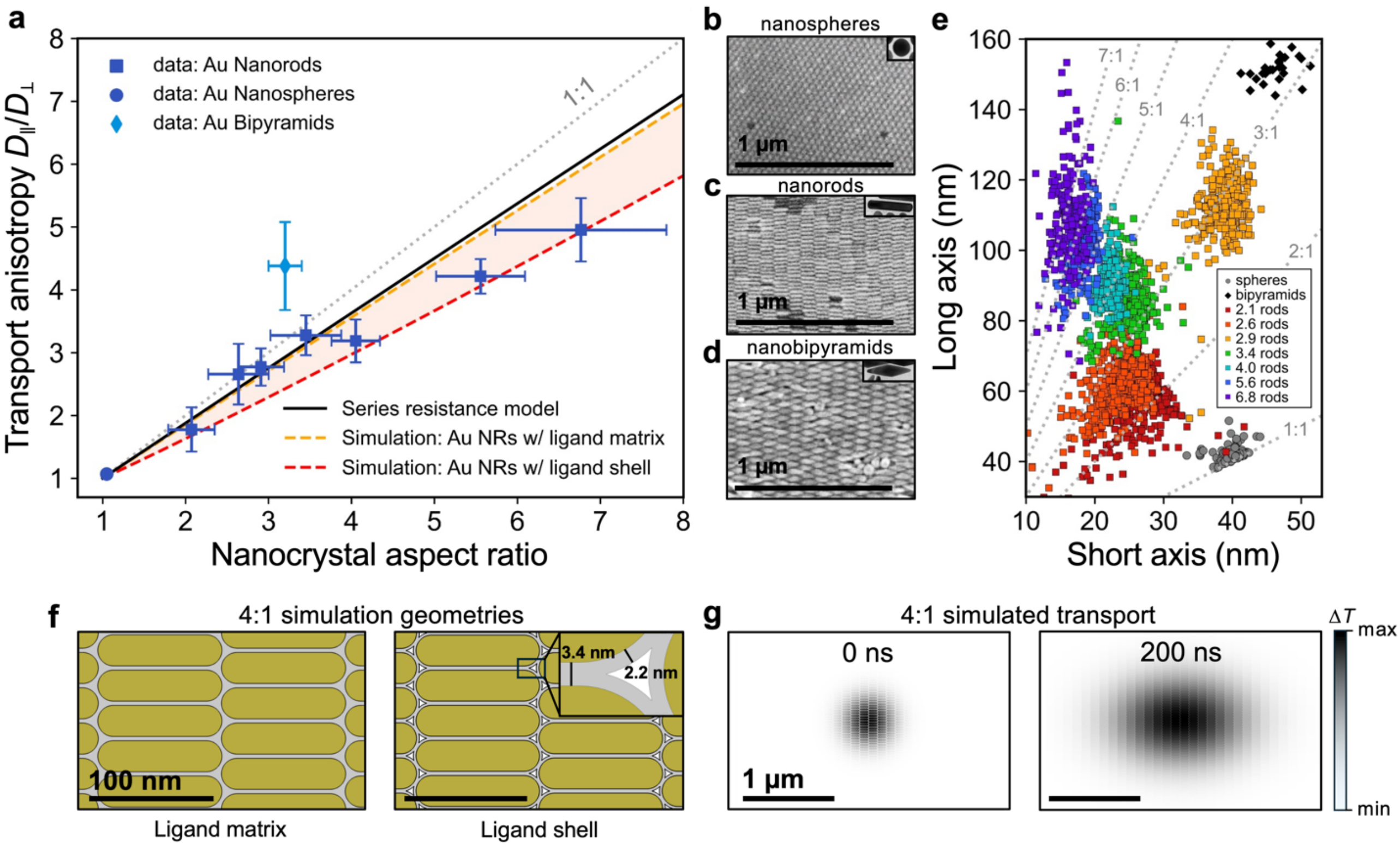

**Figure 3. Tuning and modeling thermal transport anisotropy.** **a**, Thermal transport anisotropy as a function of NC aspect ratio. Error bars on $D_{\parallel}/D_{\perp}$ are the standard error of multiple measurements on different locations while the error bars on the aspect ratio come from the standard deviations of the size distributions. The data are in reasonable agreement with the analytical series resistance model of equation (S5) and finite element simulations for the nanorod (NR) geometries shown in panel (**f**). **b-d**, Representative SEM images of supercrystals of (**b**) 41 nm diameter Au–polystyrene thiol nanospheres, (**c**) 4.0:1 (22 nm × 91 nm) Au–CTAB nanorods, and (**d**) 48 nm × 153 nm Au–CTAC bipyramids. Insets: Representative TEM images of the corresponding NCs, not to scale. **e**, Scatter plot summarizing TEM measurements of NC dimensions. Reference lines of hypothetical integer aspect ratios appear as dashed gray lines. Mean and standard deviation statistics appear in Table S2. **f**, Representative finite element simulation geometry for Au nanorods with 3.4 nm inter-NC spacing both with 2.2 nm thick CTAB ligand shell that gives small voids between nanorod tips and for the same packing but with a uniform CTAB ligand matrix. **g**, Representative simulated temperature map evolution over time for 4:1 (100 nm × 25 nm) nanorods with ligand shells, starting from a 420 nm fwhm Gaussian profile in the Au nanorod cores.

## Simulations and modeling of thermal anisotropy

In order to develop a physical picture for how nanoscale structure and thermal properties govern anisotropy, we turn to finite element simulations of heat transport. As a model geometry (Figure 3f), we construct a series of two-dimensional SCs of nanorods with representative diameters of 25 nm and rounded tips arranged in the side-by-side centered rectangular structure seen in SEM images (Figure 3c). Taking a shell of 2.2 nm long ligands with partial interdigitation based on previous reports[22] induces interstitial, nonconductive voids between the tips of the nanorods that primarily hinder transport along the major axis (Figure 3f).[19] We also consider the case of a continuous ligand matrix filling the interstices to account for the possibility of excess ligand in the SCs.[38] We use literature values for the thermal conductivity, density and heat capacity of the gold and ligand matrix, and explicitly include the interfacial thermal conductance of the NC–ligand interface (see Methods on finite element simulations). Given these simulation geometries, we approximate the spatiotemporally-resolved thermoreflectance experiments by starting from an initial isotropic Gaussian temperature profile and compute heat propagation in space and time according to Fourier's law (Figure 3g, S7, Table S5). The simulations reproduce the anisotropic thermal transport seen in experiments with the data falling between the two model geometry results within the measurement uncertainty (Figure 3a) and give reasonable quantitative agreement with the measured diffusivities (Table S3). We emphasize that we did not perform parameter fitting to achieve such agreement, but rather used reasonable values from the literature. Our findings show that the experimental observation of anisotropic thermal transport can be accounted for simply by invoking Fourier's law, reasonable thermal parameters, and an anisotropic composite material geometry. The three-dimensional arrangement of deeper SC layers could modulate the anisotropy, but the general principles established here could be used to treat such specific cases.

To establish a set of design principles regarding the material parameters and structure that govern thermal transport anisotropy in NC SCs, we explore an analytical equivalent thermal circuit model. Similar effective medium approximation models have been successfully employed to predict the thermal conductivity of isotropic assemblies of spherical NCs.[10–13,39] Here, we consider a simplified geometry of two independent one-dimensional series resistances for the major and minor axes of the SC (Figure S8, S9) to obtain the thermal diffusivity ratio (or thermal conductivity ratio) as a function of the NC and ligand dimensions and thermal properties (equation (S5)). A plot of $D_{\parallel}/D_{\perp}$ as a function of nanorod length-to-diameter aspect ratio $L/d$ using the same dimensions and literature thermal properties as used in the simulations suggests that this simple model captures the essential physics of side-by-side close-packed nanorods in a continuous ligand matrix (Figure 3a, S10). In this picture, the thermal conductivity is higher along the major axis because the effective NC volume fraction is larger along that axis. The anisotropy can consequently be tuned through NC dimensions, increasing with aspect ratio. For a given aspect ratio, the anisotropy is geometrically limited to $D_{\parallel}/D_{\perp} \leq L/d$ (note that this limit is exceeded in other NC packing geometries as discussed below). However, the larger the NC or the shorter the inter-NC gap, the higher the maximum achievable thermal anisotropy, tending toward the aspect ratio itself. A second approach to control the thermal anisotropy is through the constituent material thermal properties. To achieve maximum anisotropy for a given NC SC geometry, one could maximize the NC core thermal conductivity with respect to either that of the ligand or the interfacial thermal conductivity (Figure S11, S12), which is effectively achieved already in the present Au NCs (Figure S13). However, in the case of colloidal nanorod SCs with typical organic ligands, the conductivity contrast between the NC core and the ligands is more important for the thermal anisotropy because common interfacial conductances[10,11,13,15] are relatively large (Figure S11, S12, S13). Interestingly, while the thermal conductivity is dominated by the ligand,[10] the anisotropy—for a given aspect ratio and ligand or NC size—is governed by the thermal conductivity contrast. As discussed below, the NC shape and packing arrangement also controls anisotropy in ways not captured by this analytical model, but we expect these general strategies to increase anisotropy to hold.

**Beyond aspect ratio**

We now pose the question of whether the thermal transport anisotropy can be increased above the apparent limit imposed by the NC aspect ratio by exploring the NC shape and lattice symmetry. In a spatiotemporal picture, heat propagation undergoes a series of steps transferring slowly across ligand barriers then rapidly across the extent of the anisotropic NC akin to asymmetric hopping steps bottle-necked by the ligand. While parallel and perpendicular heat transport are nearly independent in the above side-by-side nanorod SC packing structure, NC packing geometries that couple transport along these axes could leverage this rapid intra-NC heat transport by repurposing, through diagonal paths, small advances along the slow-axis to make additional contributions along the fast-axis bypassing the ligand barriers (Figure S14, S15, S16). We demonstrate this phenomenon in SCs of colloidal Au nano-bipyramids (Figure 3d). Measurements (Figure 3a), confirmed by finite element simulations (Figure S14), demonstrate that this particular NC geometry leads to an anisotropy ($D_{\parallel}/D_{\perp} = 4.4$) substantially greater than the NC aspect ratio (3.2:1). Similarly, nanorod assembly arrangements that couple major and minor axis transport, such as a brick wall-like pattern of end-to-end packing with staggered alignment of adjacent rows (Figure S15),[23] should also achieve $D_{\parallel}/D_{\perp} > L/d$. While the effective medium model does not account for such heat flow pathways, finite element simulations and a limiting hopping diffusion model allow for such complex geometries (see Supporting Information for further discussion). These models confirm this phenomenon and predict that it approaches a quadratic dependence on aspect ratio (Figure S16).

Finally, we explore the possibility to selectively decrease $D_{\perp}$ by reducing inter-NC coupling along one axis, similar to layered materials and epitaxial nanostructuring.[40,41] Examination of nanorod packing in SEM images here (Figure S17) and of previous reports[22,23] reveals instances in which the inter-NC spacing is larger than twice the ligand length, suggesting the presence of nanoscale voids that would hinder minor-axis thermal transport. One might expect self-assembly dynamics to lead to more complete, close-packed structures near the bottom layers facilitated by the flat SC–substrate interface and a higher frequency of incomplete, non-close-packed NC arrangements near the top layers. With this picture in mind, together with the notion that the optical penetration depth (see Methods on attenuation length) makes the reflectance-based method weighted by the surface layers of our multilayer SCs, we compared the thermal transport anisotropy of 4.0:1 and 5.6:1 nanorod SCs when probed from below (through the coverslip substrate) and from above (by flipping the sample over). In contrast to the measurements made from below presented in Figure 3a that yielded $D_{\|}/D_{\perp} < L/d$ with relatively small standard error, we obtained consistently high anisotropies in the range of 5–16 when probing the top layers (Figure S18, Table S6). A wide range of measured anisotropies in both samples (Figure S18) indicates a large degree of sample heterogeneity for the top layers, suggesting a strong sensitivity to the precise size and arrangement of interstitial voids. We confirm by finite element simulations that non-conductive voids between adjacent nanorods lead to anisotropies greatly exceeding the aspect ratio by hindering transport along the minor axis (Figure S19). We note that while this particular strategy implies less synthetic control than tuning NC shape at this time, existing protocols for asymmetric surface functionalization could potentially turn this phenomenon into a rational material design.[42,43] However, compared to increasing anisotropy by tuning NC shape and packing arrangement, introducing voids may increase anisotropy at the expense of poorer overall heat dissipation (Table S5 and S6), which may be an important consideration for some applications.

## Discussion

We expect significant, tunable thermal transport anisotropy to be a general phenomenon in self-assembled NC SCs for different material compositions and size. As shown previously, the thermal conductivity of NC films is nearly independent of NC core material (including various semiconductor, metal oxide and metallic NCs) because slow transport across the interstitial ligand matrix and the NC–ligand interface dominate. Analogously, we calculate that thermal anisotropy should persist in most common NC core materials and sizes (Figure S13). However, there is a trade-off in that lower NC core thermal conductivity and smaller NC size reduce the anisotropy (Figure S11, S12). Similarly, the series resistance model predicts strong thermal anisotropy across the range of common ligand lengths, approaching the NC aspect ratio when going from long, insulating aliphatic ligands to short ligands that also increase electronic coupling[6] and overall thermal conductivity[10,13] (Figure S13). We also point out that while ligands were shown to dominate the thermal conductivity of NC solids, changing the identity of the ligand only leads to minor quantitative differences,[10,13] and our calculations further show that the anisotropy is robust to moderate changes in the magnitude of the ligand thermal diffusivity (Figure S11, S12, S13). For example, interchanging the different ligands used here does not significantly change the expected results (Table S7). In the hypothetical limit of removing the ligand entirely, the thermal anisotropy would still approach the geometric limits if a substantial interfacial resistance were maintained between the original NC cores.

We envision that the principles established here can be applied to design anisotropic heat transport for a wide range of SC packing structures and NC shapes. Assemblies of anisotropic NCs have a richer phase behavior than explored here,[1,2,22,23] which could be used as another layer of control for nanoscale thermal anisotropy. Similarly, higher orders of hierarchical structure and directed, templated and patterned self-assembly could be used to design guided mesoscale heat flow.[1,2,44,45] Our observation of heat following the nanorod grain orientations across tilt grain boundaries and curved arrangements in Figure 2 motivates the design of heat flow along non-straight paths within a two- or three-dimensional material. Beyond one-dimensional building blocks, two-dimensional nanocrystals such as nanoplatelets could be used in analogy to layered materials to induce two high-conductivity dimensions instead of one.

Tunable thermal transport anisotropy presents opportunities for controlling heat flow in the various applications that use NCs such as optoelectronics, thermoelectrics, phase-change memory, nanoelectronics and catalysis. Thermal management is an important design feature in devices as heating due to optical excitation or current leads to effects such as performance degradation or instability, losses,[46] and unwanted chemical activity.[47] Past strategies include improving NC thermal stability,[48] implementing heat sinks and active cooling,[49] and using intermittent duty cycles.[50] Leveraging anisotropic thermal transport presents an additional strategy to enhance thermal dissipation in such devices while maintaining control over size-dependent optical and electronic effects. Anisotropic thermal dissipation may already be unknowingly at play in applications that use ordered nanorod arrays, such as plasmonics, polarized lighting and display, and photoconductors.[1,51–54] Beyond enhancing thermal dissipation, the behavior uncovered here poses the intriguing possibility to intrinsically funnel heat to or away from desired locations within a device architecture using the device's active material without heat sinks. In addition to thermal management, tunable thermal anisotropy poses interesting prospects for thermoelectric applications that rely on control over thermal gradients and for phonon engineering.[18,55] For all of the above, self-assembly of solution-processed NCs proposes a cost-effective alternative compared to epitaxial or lithographic nanostructuring-induced anisotropy and with a comparable magnitude of anisotropy.[41] Furthermore, by virtue of using colloidal NC building blocks, the optical and electronic properties of the active material itself can be tuned in addition to the thermal properties.

## Conclusions

Our work demonstrates colloidal NC self-assembly as a route to engineering anisotropic thermal transport. Modeling combined with correlative electron microscopy and spatiotemporal optical measurements identify macroscopic anisotropy as an emergent property generated by tunable nanoscale building blocks. Perhaps counterintuitively, the anisotropy of the nanoscale constituents is not erased—rather is maximized—by the interfaces and ligands. This phenomenon may be used to realize solution-processed devices with directed heat transport along specific axes with a three-dimensional active material. Model predictions taken together with the well-explored modular nature of NC solids imply the ability to implement thermal anisotropy in metamaterials with a wide range of tunable optical, electronic, mechanical and catalytic properties. We expect correlative SEM and pump–probe microscopy to be a powerful approach for revealing structure-function relationships in nanomaterials more generally.

## METHODS

### Materials

Hydrogen tetrachloroaurate trihydrate ($HAuCl_4$•$3H_2O$, ≥99.9%, Sigma-Aldrich USA), silver nitrate ($AgNO_3$, 99%, Johnson Matthey, UK), trisodium citrate dihydrate ($C_6H_5Na_3O_7$•$2H_2O$, ≥99.9%, Sigma-Aldrich USA), sodium borohydride ($NaBH_4$, 99%, Acros Organics), cetyltrimethylammonium bromide and chloride (CTAB and CTAC, ≥98% Sigma-Aldrich, USA), benzyldimethylhexadecylammonium chloride (BDAC, 97%, Sigma-Aldrich, USA), sodium oleate (NaOL, 97%, TCI, Japan), L-ascorbic acid (AA, 99%, Sigma-Aldrich, USA), and hydrochloric acid (HCl, 37 wt. % in water, Carlo Erba, Italy), toluene (≥99.5%), tetrahydrofuran (≥99.5%), and ethanol (denat., >96%, VWR, USA), DEG (reagent grade, Merck, Germany) and thiolated polystyrene (PSSH, $M_n = 5300$ g mol$^{-1}$, $M_w = 5800$ g mol$^{-1}$, Polymer Source, Canada) were used as received. Ultrapure water (resistivity: 18.2 MΩ•cm) was used for all preparations.

### Synthesis, functionalization and assembly of Au nanorods

The Au nanorods were synthesized according to a previous report.[56] The nanorod seeds were prepared first. 364 mg of CTAB were dissolved in 9.75 mL of water in a 50 mL vial. 250 µL of 10 mM $HAuCl_4$ were then added and the solution took on a yellow/orange color. 600 µL of freshly made 10 mM $NaBH_4$ was then added rapidly under vigorous stirring. The seed solution was vigorously stirred for 2 minutes then the stirring was stopped and the solution was kept undisturbed for 30 minutes before being added to the growth solution. At this point the solution was brown. The seed solution kept at 28 °C would be stable for a couple days before turning pink/red. To grow the nanorods, 360 mg of CTAB and 45 mg of NaOL were dissolved in 18 mL of water in a 50 mL plastic vial. Dissolution was complete after leaving the closed vial in a 50 °C water bath for a few minutes, followed by gentle hand stirring. The mixture was then cooled down in a 30 °C water bath. Freshly made 4 mM $AgNO_3$ solution was then added in varying amounts (Table S1). The solution was kept undisturbed for 15 minutes. Then 1 mL of 10 mM $HAuCl_4$ was added and the solution was left under medium-speed stirring for 90 minutes. The solution took on a yellow/orange color and turned transparent after about 15 minutes due to reduction of $Au^{3+}$ to $Au^{+}$ by NaOL. 37% HCl was then added under slow stirring for 15 minutes (Table S1). 50 µL of 64 mM ascorbic acid was added under strong stirring for 30 seconds, followed by the addition of 20 µL of the seed solution under strong stirring for 30 seconds. The solution was then left undisturbed for 15 hours at 30 °C. The solution was then centrifuged at 5500 g for 45 minutes and the supernatant was carefully removed. The precipitate was redispersed in 3 mL of 2.5 mM CTAB and centrifuged at 5500 g for 20 minutes. The supernatant was removed and replaced by 1.5 mL of 1 mM CTAB. This solution is referred to as the nanorod stock solution below. In order to obtain a highly concentrated Au nanorod solution for supercrystal self-assembly, 500 µL of the stock solution was centrifuged once more and redispersed in 100 µL of 1 mM CTAB.

The Au nanorods whose aspect ratio was 6.8:1 had a poor shape yield (around 40%) were purified by flocculation.[57] 600 µL of the Au nanorod stock solution described above were mixed with 300 µL of BDAC (480 mM) in a 2 mL Eppendorf tube and left undisturbed for 1 hour in a 35 °C oven. The purple/pink supernatant (containing mostly nanocube and nanosphere by-products) was transferred to another Eppendorf tube and the black precipitate (containing mostly Au nanorods) was redispersed in 1 mL of 10 mM CTAB. The purified Au nanorods were centrifuged at 12500 g for 3 min and redispersed in 1 mL of 2.5 mM CTAB. The purified Au nanorods were then centrifuged again at 12500 g for 3 min and redispersed in 500 µL of 1 mM CTAB. Our estimation is that the morphology yield of Au nanorods was more than 90% after purification. The Au nanorod batch with an aspect ratio of 4.0:1 used for correlative SEM and stroboSCAT measurements was also purified the same way.

To self-assemble nanorod supercrystals, we use a well-established protocol that reliably gives large, several-micron domains.[23] Here, 3 µL of a concentrated solution was drop-casted on a clean glass coverslip or polished (100) Si wafer. 20 mL of hot tap water (~70 °C) were added to a small beaker and placed beside the substrate. The whole system was then covered by a crystallizer and the drop was left to dry overnight in this water vapor-saturated environment. The drying taking place in this humid atmosphere takes around 8 hours, compared to the drying of a 3 µL droplet in normal air which takes around 1 hour.

**Synthesis, functionalization and assembly of Au pentagonal bipyramids**

The synthesis of the Au pentagonal bipyramids were carried out following a recent article by seed mediated growth from using penta-twinned seeds.[24] In a typical synthesis of the penta-twinned seeds, 2.645 mL of 25 wt% CTAC, 32.96 mL of water, 0.4 mL $HAuCl_4$ (25 mM) were taken in a 100 mL Erlenmeyer flask and stirred gently at 30 °C for 10 min. 4 mL of trisodium citrate dihydrate (50 mM) was added to the above mixture and the solution was kept at 30 °C for 30 min followed by addition of 1 mL of freshly prepared $NaBH_4$ (25 mM). The mixture was stirred vigorously for 30 seconds and kept for 5 days in the oven at 40 °C without stirring for aging. These nanoparticles were used as seeds for the synthesis of Au bipyramids. In a 500 mL Erlenmeyer flask, 100 mL of 100 mM CTAB, 1 mL of 10 mM $AgNO_3$, 2 mL of 25 mM $HAuCl_4$, 2 mL of 1 M HCl and 0.8 mL of 100 mM Ascorbic acid were mixed by gentle stirring at 30 °C. 0.125 mL pentatwinned seeds were added into the above mixture and the solution were kept at 30 ℃ for 4 hours. The as-synthesized Au bipyramids were collected by centrifugation at 3000 RPM. Purification of the bipyramids was carried out by depletion using 250 mM BDAC. The particles were washed with 2.5 mM CTAC (25 wt % in $H_2O$, 90%) and stored in 2.5 mM CTAC for further use.

Au bipyramids supercrystals were prepared by evaporation-induced self-assembly using a glass cover slip (20 mm × 20 mm, thickness 175 µm) or Si-wafer (cleaned by washing with water, ethanol and acetone). A 10 µL solution of the nanocrystals was dropcasted on the substrate and dried in a covered petri-dish.

**Synthesis, functionalization and assembly of Au nanospheres**

The Au nanospheres where synthesized and assembled following established protocols.[58,59] In brief, the synthesis is a seeded-growth protocol with cetyltrimethylammonium chloride as a capping ligand, which is then exchanged with thiol-terminated polystyrene-ligands (here with an average molecular weight of ~5000 g $mol^{-1}$) by phase-transfer into tetrahydrofuran and finally into toluene. The purified dispersions of polystyrene-capped Au nanospheres in toluene are then pipetted onto a liquid subphase (diethylene glycol) in a teflon well. Slow evaporation of the toluene then leads to well-ordered self-assembly of the Au nanospheres in the thin-film supercrystals with layer numbers ranging from mono- to multilayers. A detailed description of the synthesis and characterization of the materials can be found in Ref 58.

Here we report thermal transport measurements performed on multilayer domains of nanosphere supercrystals.

**Transmission electron microscopy (TEM)**

For TEM of the Au nanorods, 3 µL of stock solution were deposited on a carbon coated TEM copper grid. TEM images were recorded with a JEOL 1011 at 100 kV. Images were analyzed with Image-J. At least 5 images with 40k magnification were used for size distribution measurements.

TEM for the Au bipyramid nanocrystals were measured on a copper grid (200 mesh) using a JEOL 1400 operated at 120 kV, located at Institute for Integrative Biology of the Cell, University Paris-Saclay, France.

Samples of self-assembled Au nanosphere supercrystals were carefully skimmed off the liquid subphase with a carbon-coated copper grid (400 mesh) held by a tweezer. The grid was then dried in vacuum for at least 1 h. These samples were also measured with a JEOL 1011 at 100 kV and the measurements analyzed with Image-J.

**Scanning electron microscopy (SEM)**

The Au nanorod supercrystals were imaged using a Hitachi SU-70 SEM-FEG operated with a 5 kV acceleration voltage at the platform of Institut des Matériaux de Paris-Centre, Paris, France.

The Au bipyramid supercrystals were imaged using a SEM-Zeiss Supra55VP at Laboratoire de Physique des Solides (LPS, Orsay, France) with 3 kV acceleration voltage.

Au nanosphere assemblies were measured with a Zeiss Sigma at 10 kV.

**UV-visible absorption spectroscopy**

Absorption spectra of the Au nanorods were recorded using a Cary-5000 spectrometer. Colloidal solutions were prepared by diluting a known amount of a stock solution (typically 5 µL) in 1 mL of CTAB (1 mM). The spectra were recorded in a quartz cuvette with a path length of 1 cm.

The extinction spectrum of Au bipyramids was measured on a Cary 5000 UV-Vis-NIR spectrometer in disposable polystyrene cuvettes (VWR European Cat. NO. 634-0675, path length of 1 cm).

The absorption spectrum of the Au nanosphere supercrystals was measured using a Maya 2000 Pro spectrometer coupled to an upright optical microscope (Leica DMRX). The signal from a 40 µm × 40 µm region of interest was collected through an air objective (×50, NA = 0.85). The region measured consisted of a range of supercrystal layers between one and ten layers.

**Atomic force microscopy (AFM)**

AFM of the same nanorod supercrystal domain measured by modulated thermoreflectance was performed using an NX20 AFM (Park Systems) equipped with a PPP-NCHR 10M non-contact cantilever (Park Systems). Images were collected in non-contact mode. Images were flattened and corrected using Gwyddion.[60] Film thickness was determined by comparing the height relative to the glass substrate.

**stroboSCAT measurements**

The experimental setup for stroboSCAT measurements was built based on a previously described instrument developed by Delor et al.[27] The home-built microscope was an inverted microscope featuring a high numerical aperture (1.4 NA) oil-immersion objective (Leica HC PL APO 63×/1.40 NA). Samples were mounted in a microscope slide format holder and controlled by a three-axis piezo-controlled nanopositioning stage (BIO3, PiezoConcept). Laser diodes were used for the pump (LDH-D-C-405, PicoQuant) and the probe (LDH-D-C-640, PicoQuant) beams with center wavelengths of 405 and 635 nm, respectively. The pulse durations of these two lasers are both <100 ps. The laser diode heads were controlled by the PDL 828-S "SEPIA II" laser driver equipped with two SLM 828 modules and a SOM 828-D oscillator (PicoQuant). We used a laser repetition rate of 500 kHz, with the pump modulated at 660 Hz. The pump−probe delay times were controlled electronically using the laser driver with <20 ps resolution. Both pump and probe were spatially filtered through 25 μm pinholes (P25K, Thorlabs). The pump beam was expanded to give a ∼6 mm diameter collimated beam while the probe beam was reduced to ~1 mm and then focused into the back focal plane of the objective lens using an $f$ = 300 mm wide-field lens (AC508-300-AB-ML, Thorlabs). The two beams were overlapped using a long-pass dichroic mirror (DMLP505, Thorlabs), and a 50/50 beamsplitter (BSW10, Thorlabs) reflected both beams into the objective and onto the sample, resulting in an overlapped near-diffraction-limited pump and wide-field probe illumination. Probe light reflected from the sample was collected through the same objective, isolated with a band-pass filter (FBH640-10, Thorlabs), and focused onto a charged metal oxide semiconductor (CMOS) camera with 4.5 μm square pixels (PixeLINK PL-D753MU-T, equipped with a Sony IMX 421 global shutter sensor) triggered at 660 Hz using an $f$ = 500 mm lens placed one tube length (200 mm) away from the back focal plane of the objective. The total magnification is approximately 63×500/200 = 157.5, giving approximately ~29 nm/pixel. This calibration was verified using a 1951 USAF Pattern resolution target (R3L1S4PR, Thorlabs), giving 28.3 nm/pixel. stroboSCAT images are generated by taking the difference between pump-on and pump-off raw pixel intensities, normalized to the raw pump-off intensities, yielding $\Delta R/R$ contrast images. Pump-on and pump-off images were distinguished by picking off a small fraction of the pump beam and directing it onto the corner of the camera region of interest but not overlapping the pump–probe image. Averaged pump-off images are simultaneously recorded at each time delay.

Experiments were conducted at 292 K. The pump spot size was measured by imaging the reflection from a clean glass coverslip, showing a Gaussian profile with a FWHM of 340 nm. The 405 nm pump light pulses excite the d–s interband transition of the Au nanocrystals, which rapidly thermalizes with the lattice via electron–phonon scattering on a time scale faster than the ~100 ps instrument response function.[14] We note that, unlike typical thermoreflectance experiments, no transducer is needed in the present case as the heat pulse can be generated directly within the metallic NC cores, avoiding the need to account for the transducer–to–sample heat transfer timescale that could inhibit the separation of timescales needed to extract the supercrystal thermal diffusivity. The Au NC films were excited with a $2\sigma$-integrated fluence of ≈50 μJ/cm$^2$, corresponding to an approximate temperature rise of a few Kelvin after lattice equilibration (see estimation of absorption length below).[14] We verified that the measured diffusivities did not depend on pump fluence over the range of 30–100 μJ/cm$^2$, and we ensured that no sign of film damage was observed at the pump fluence used. We image the resulting temperature-induced $\Delta R/R$ spatial profile in wide-field using 640 nm probe light pulses for pump–probe time delays typically out to 200 ns. Based on the common assumption that $\Delta R/R$ profile is proportional to the transient temperature change of Au, the lateral expansion of this profile directly reflects the thermal transport in the SC. We focus on the diffusive regime corresponding to the effective diffusion coefficient of the SC composite medium, after the timescale of inter-NC heat transfer. In this system we measure root-mean-squared displacements between ~30–400 nm (Figure S6), meaning that regions of size less than even 1 μm are sufficient to probe thermal transport properties in an area-selective manner.

Data acquisition was implemented in LabVIEW 2022 64-bit. Data analysis and plotting were performed using a combination of ImageJ, Igor Pro, and Python. To obtain the major and minor axis mean-squared expansion curves, each $\Delta R/R$ image of a given stroboSCAT time series is fitted to a two-dimensional elliptical Gaussian function with a common center and major axis angle for all times and floating amplitude and major- and minor-axis widths. The optimal elliptical Gaussian centers and orientation angles are found by first minimizing the sum of the squared residuals using the Levenberg-Marquardt algorithm for each time point separately, then in a second round these values are used as interval bounds to fit common centers and angles using the Trust Region Reflective algorithm. We repeat this process iteratively until the center and orientation angle constraint intervals converge to 5 nm and 1°, respectively. The diffusion coefficients are then obtained from the mean squared expansion curves as described in the main text.

**Optical microscopy**

Optical microscopy images were collected on the stroboSCAT setup using a white-light light emitting diode (LED) (MNWHL4, ThorLabs) that was equipped with an adjustable collimating lens (SM1U25-A, ThorLabs) and driven by a LED driver (T-Cube LEDD1B, ThorLabs).

**Estimation of nanocrystal supercrystal absorption length and temperature jump**

The absorption length of a SC of 4:1 nanorods was estimated by measuring the absorption of a focused laser beam through a sample of known thickness. With the sample mounted on a glass coverslip on the stroboSCAT microscope, a focused 405 nm beam used for the pump source was positioned within an optically-smooth region of a SC domain. The power passing through the SC relative to the power passing through a blank region of coverslip was measured using a S171C silicon photodiode power sensor and PM100D power meter (Thorlabs). The optical density was found to be $A = 1.9$. The thickness of the same location was measured using AFM (Figure S4) to be $l = 550$ nm. The attenuation length is then defined as $\alpha^{-1} = l/A$, giving 290 nm at a wavelength of 405 nm. While the absorption length is expected to change with NC size and thus aspect ratio, we take this sample to be representative of the order of magnitude for the depth penetrated in our experiments.

The pump-induced lattice temperature jump can be estimated as $\Delta T = U/C_L$,[14] where $U$ is the absorbed energy per unit volume and $C_L$ is the volumetric heat capacity. We estimate $U$ based on the pump fluence and absorption length, and we assume the value of bulk Au of $C_L = 2.5 \times 10^6$ J m$^{-3}$ K$^{-1}$, giving $\Delta T \approx 1$ K. The exact value for each sample depends on the size of the nanorod, but will be in this range $\Delta T \sim 1$ K. Performing this estimation using literature values[61] for the optical extinction coefficient at the plasmon resonance of Au nanorods combined with our absorption spectra gives a similar value.

**Frequency domain modulated thermoreflectance microscopy**

Thermal transport on a 10 μm length scale was evaluated using modulated thermoreflectance (MTR) microscopy. In this experiment, the pump beam (532 nm, Cobolt MLD) was raster scanned across the sample and absorbed directly by the nanocrystal film while the resulting change in reflectance was probed using a fixed probe beam (488 nm, Oxxius). The beam spot size is approximately 1.5 μm. The intensity of the pump beam was modulated by an acousto-optic modulator at 100 kHz and the beam was focused on the sample with an objective lens (N.A. = 0.5). The change in reflectance of the probe beam was measured using a photodiode and a lock-in amplifier to record the AC reflectivity component. A pump beam power of 70 μW and probe beam power of 80 μW were used. The experimental amplitude and phase profiles were fit according to numerical simulations based on Fourier's law to extract the thermal transport anisotropy of the films using a 1.5 μm Gaussian pump beam diameter and diffraction-limited Airy probe profile.[62,63] A detailed procedure of the data treatment can be found from ref 64.

**Finite element simulations**

COMSOL Multiphysics® 6.0 (COMSOL Inc., Los Angeles, CA) finite element software was used to simulate heat transport in the Au NC SCs with the Heat Transfer in Solids module. To render the simulations computationally feasible, we reduce the geometry to two dimensions and study the in-plane thermal transport. Simulation geometries were constructed with packing structures based on SEM images. The nanorod aspect ratios are varied from with a fixed diameter of 25 nm as a representative size of the nanorods studied here. The nanorod shapes account for the rounded tips seen in TEM images (Figure S3) and are surrounded with a well-defined 2.2 nm thick ligand shell. We choose an inter-NC spacing of 3.4 nm based on previous reports of Au nanorods with CTAB ligands,[22] which includes partial ligand interdigitation. This geometry has the particularity to induce interstitial voids between the tips of the nanorods (Figure 3f) that we take to be nonconductive (Figure 2f,g). We also consider the case of a continuous ligand matrix filling the interstices (Figure 3f) to account for the possibility of smaller spacing or excess ligand in the un-washed SCs.[38] The lateral sizes of the simulation geometries were large enough to avoid edge effects in the temperature gradient evolution.

Au NC cores were given the properties of bulk Au: density of 19,300 kg $m^{-3}$, thermal conductivity of 310 W $m^{-1}$ $K^{-1}$, and heat capacity of 129 J $kg^{-1}$ $K^{-1}$. The Au nanorod and nano-bipyramid samples had CTAB and CTAC ligands, respectively, which we assumed to have the same thermal properties based on literature values: density of 500 kg $m^{-3}$,[43] thermal conductivity of 0.15 W $m^{-1}$ $K^{-1}$,[65] and heat capacity of 1400 J $kg^{-1}$ $K^{-1}$.[66] The Au–CTAB/CTAC interface is treated as an equivalent thin resistive layer assuming an interfacial thermal conductance of 40 MW $m^{-2}$ $K^{-1}$ based on previous reports for Au nanocrystals with amine-based ligands.[13,67] Thermal transfer to the air (including voids) or substrate was neglected in the simulations.

300 K was used as the ambient reference temperature, and simulations were initialized using a Gaussian profile (fwhm = 420 nm) of elevated temperature with a peak of 310 K ($\Delta T_{max}$ = 10 K) restricted to the Au NC cores, approximating the experimental conditions. The mesh was constructed using a "fine" element size. The finite element simulation solves the heat equation using a quadratic Lagrange discretization to calculate the time dependence of the temperature profile with a relative tolerance of 0.0001.

The simulated temperature profiles as a function of time were treated analogously to the stroboSCAT data, fitting to a Gaussian function over time and calculating $\sigma_{\parallel,\perp}^2(t) - \sigma_{\parallel,\perp}^2(t_0)$ to produce mean-squared expansion curves along the major and minor axes and thereby extract the diffusivity by fitting to $2D_{\parallel,\perp}t$ in the diffusive time regime. The thermal transport anisotropy is calculated as $D_{\parallel}/D_{\perp}$.

## ASSOCIATED CONTENT

### Pre-print

A pre-print of this manuscript is available: Matias Feldman; Charles Vernier; Rahul Nag; Juan J. Barrios-Capuchino; Sébastien Royer; Hervé Cruguel; Emmanuelle Lacaze; Emmanuel Lhuillier; Danièle Fournier; Florian Schulz; Cyrille Hamon; Hervé Portalès; James K. Utterback. Anisotropic Thermal Transport in Tunable Self-Assembled Nanocrystal Supercrystals. 2024, arXiv:2407.08325. arXiv. https://doi.org/10.48550/arXiv.2407.08325 (accessed November 10, 2024).

### Supporting Information

Nanocrystal characterization, modulated thermoreflectance imaging of long-range anisotropy, correlative SEM and spatiotemporally-resolved thermoreflectance, spatiotemporally-resolved thermoreflectance for different nanorod aspect ratios, thermal diffusivity measurement results, estimation of thermal conductivities, finite element simulations of anisotropic thermal transport, effective medium approximation modeling of series resistance, increased thermal transport anisotropy through nanocrystal shape and packing, diffusive hopping model, increased thermal transport anisotropy by decreased side-to-side nanocrystal coupling, and comparison of different ligands.

## AUTHOR INFORMATION

### Author contributions

M.F. and J.K.U. performed spatiotemporal thermoreflectance experiments, performed simulations and modeling, analyzed results and wrote the paper with the input of all authors. C.V. and H.P. prepared and characterized the nanorod samples. R.N. and C.H. prepared and characterized the nano-bipyramid samples. J.B-C. and F.S. prepared and characterized the nanosphere samples. S.R., E.L. and E.L. contributed to experimental development. H.C. performed AFM. D.F. performed frequency domain thermoreflectance. J.K.U. conceived and directed the project.

### Competing interests

The authors declare no competing interests.

## ACKNOWLEDGMENTS

J.K.U. is grateful to Milan Delor, Raj Pandya, and Massimiliano Marangolo for useful discussions and support while getting installed at the Institut des NanoSciences de Paris, as well as to Florent Margaillan and Thierry Barisien for access to optical components, and David Montero along with the Fédération de Chimie et Matériaux de Paris-Centre (FCMat) for SEM-FEG imaging. This work was supported by startup funds from the CNRS and the Emergence Research Programme of Alliance Sorbonne Université. Financial support for this project was provided by the Institute of Materials Science (iMAT) of the Alliance Sorbonne Université. M.F. acknowledges iMAT for a PhD grant.

**SUPPORTING INFORMATION**
**for**

# Anisotropic Thermal Transport in Tunable Self-Assembled Nanocrystal Supercrystals

Matias Feldman,[1] Charles Vernier,[2] Rahul Nag,[3] Juan J. Barrios-Capuchino,[4] Sébastien Royer,[1] Hervé Cruguel,[1] Emmanuelle Lacaze,[1] Emmanuel Lhuillier,[1] Danièle Fournier,[1] Florian Schulz,[4] Cyrille Hamon,[3] Hervé Portalès,[2] James K. Utterback[1]*

[1] Sorbonne Université, CNRS, Institut des NanoSciences de Paris, 75005 Paris, France
[2] Sorbonne Université, CNRS, MONARIS, 75005 Paris, France
[3] Laboratoire de Physique des Solides, CNRS and Université Paris-Saclay, 91400 Orsay, France
[4] Institute for Nanostructure and Solid State Physics, University of Hamburg, Luruper Chaussee 149, 22761 Hamburg, Germany

*Correspondence to: james.utterback@sorbonne-universite.fr

**Table of Contents**

## SUPPLEMENTARY SECTION 1

## NANOCRYSTAL CHARACTERIZATION

### Optical microscopy images of nanocrystal supercrystals

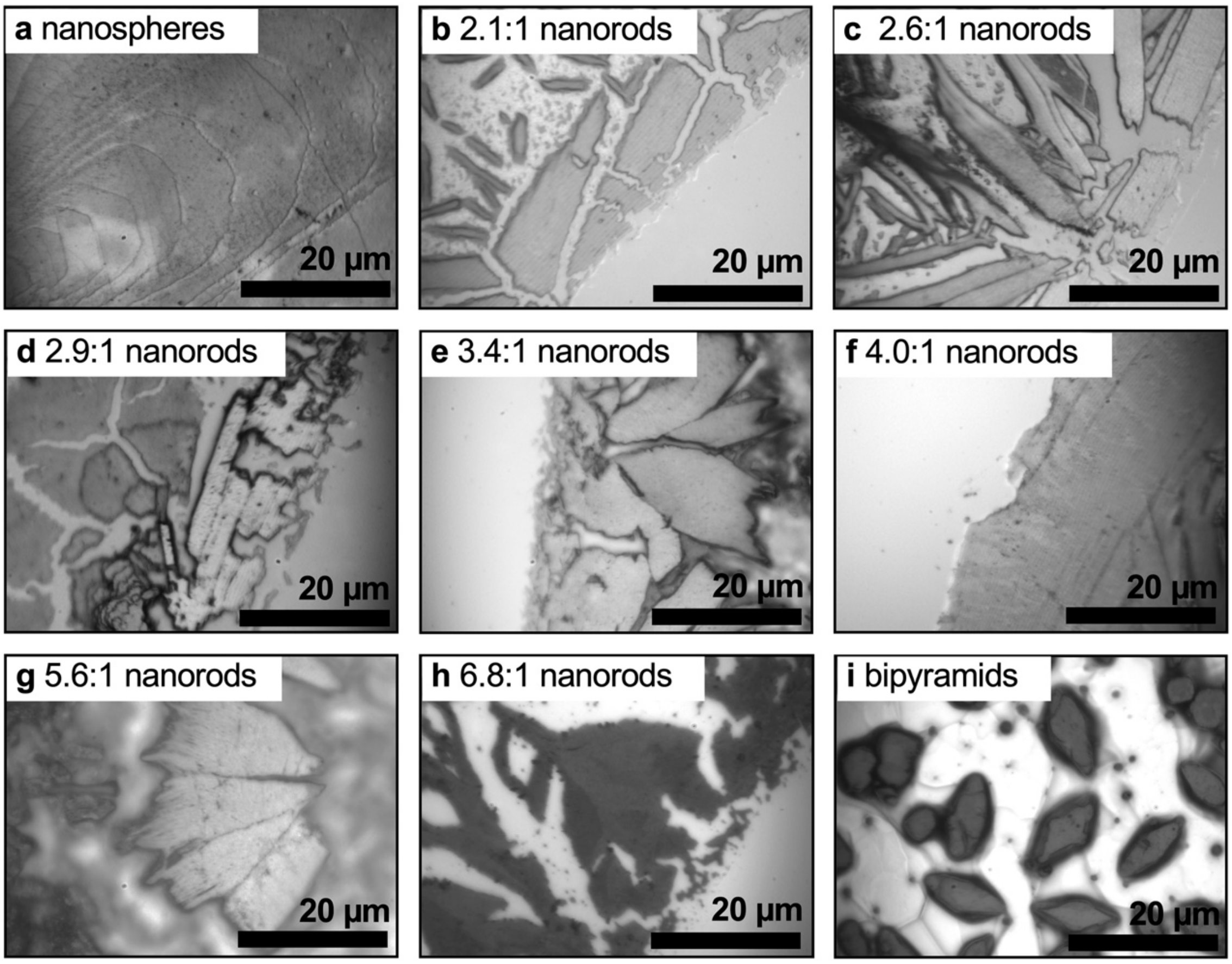

**Figure S1. Representative optical reflectance images of supercrystal domains investigated by spatiotemporal thermoreflectance measurements. a**, Multilayer supercrystal of 41 nm Au–PSSH nanospheres. **b-f**, Examples of Au–CTAB nanorod supercrystal samples used in Figure 3 with mean aspect ratios 2.1, 2.6, 2.9, 3.4, 4.0, 5.6, and 6.8. **i**, Rhomboidal supercrystals of Au–CTAC nano-bipyramids. Images were collected with a broadband LED on the spatiotemporal thermoreflectance microscope.

### Optical characterization

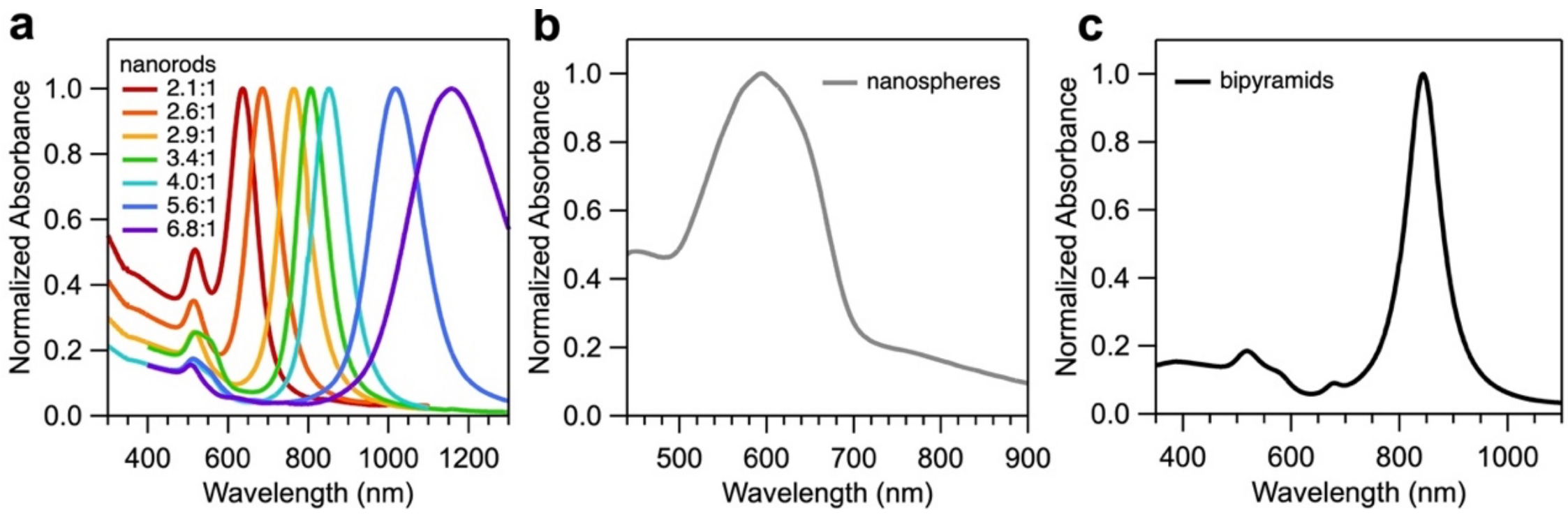

**Figure S2. UV-visible absorption spectra. a**, Au–CTAB nanorod samples in aqueous solution before assembly, normalized to plasmon resonance peak. **b**, Self-assembled Au–PSSH nanosphere supercrystal. The spectrum is collected over a region containing a range of the number of nanocrystal supercrystal layers. **c**, Au–CTAC nano-bipyramid sample dispersed in water at 2.5 mM.

## Nanorod synthesis recipes

**Table S1. Summary of Au–CTAB nanorod synthesis details**

| Aspect Ratio | CTAB (mg) | NaOL (mg) | Water (mL) | $AgNO_3$ (4 mM) | $HAuCl_4$ (10 mM) | HCl | Ascorbic acid (64 mM) | Seeds (µL) | Shape yield |
|---|---|---|---|---|---|---|---|---|---|
| 2.1 | 360 | 45 | 18 | 0.25 mL | 1 mL | 60 µL | 50 µL | 20 µL | 85% |
| 2.6 | 360 | 45 | 18 | 0.35 mL | 1 mL | 60 µL | 50 µL | 20 µL | 85% |
| 2.9 | 360 | 55 | 18 | 1.05 mL | 1 mL | 60 µL | 50 µL | 20 µL | 85% |
| 3.4 | 360 | 52 | 18 | 0.95 mL | 1 mL | 60 µL | 50 µL | 20 µL | 80% |
| 4.0 | 360 | 45 | 18 | 1.05 mL | 1 mL | 60 µL | 50 µL | 20 µL | 90% |
| 5.6 | 360 | 45 | 18 | 1.05 mL | 1 mL | 100 µL | 50 µL | 20 µL | 75% |
| 6.8 | 360 | 45 | 18 | 1.05 mL | 1 mL | 210 µL | 50 µL | 20 µL | 90% |

## TEM images of nanocrystals

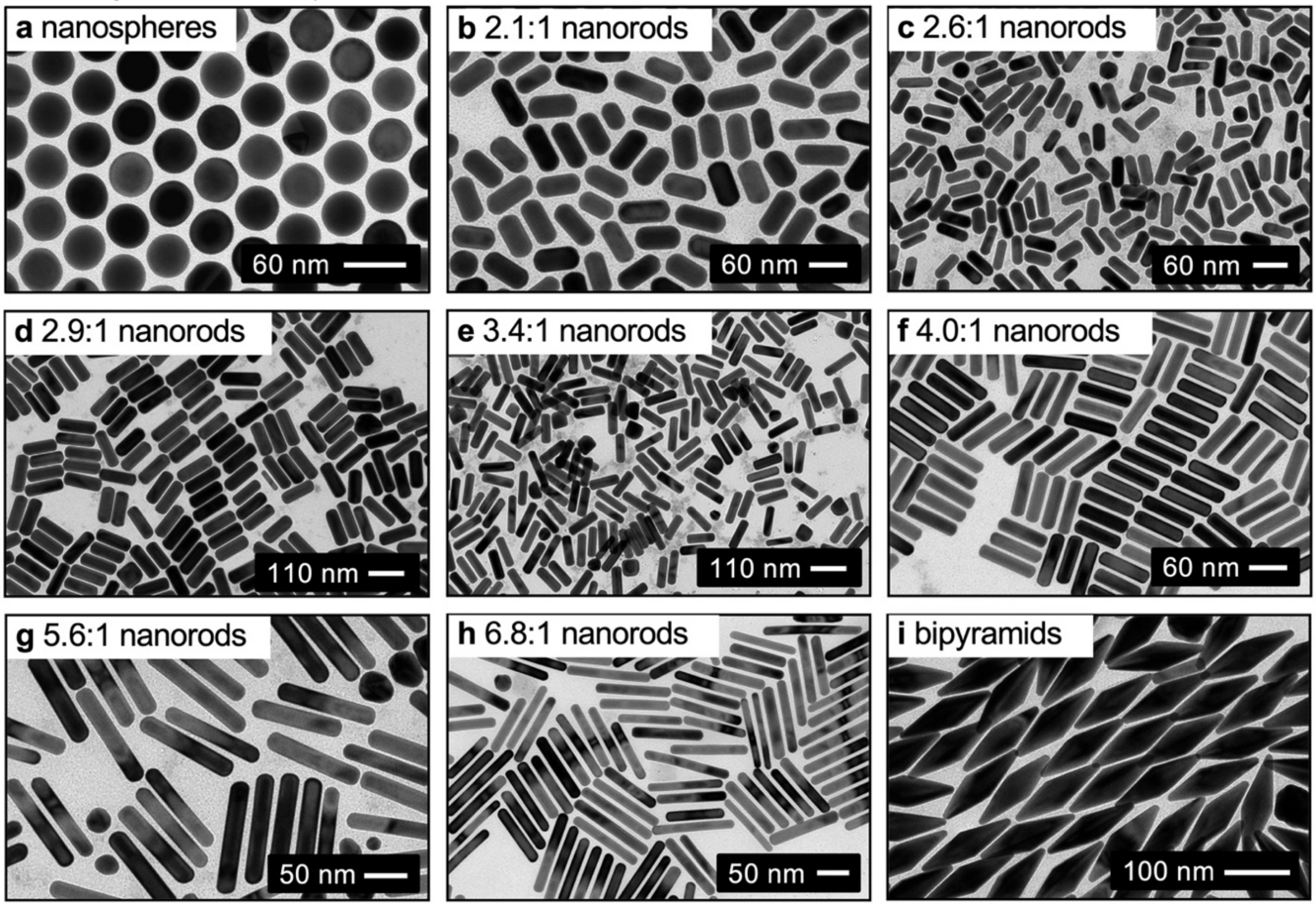

**Figure S3. Electron microscopy characterization of nanocrystal samples.** Representative TEM micrographs of **a**, Au–PSSH nanospheres; **b-h**, Au–CTAB nanorods with mean aspect ratios 2.1, 2.6, 2.9, 3.4, 4.0, 5.6, and 6.8; **i**, Au–CTAC nano-bipyramids.

**Table S2. Summary of nanocrystal size measurements**

| Sample Title | Length (nm) | Diameter (nm) | Aspect Ratio [a] |
|---|---|---|---|
| nanospheres | – | 40.6 ± 0.6 | 1.05 ± 0.05 |
| 2.1:1 nanorods | 56 ± 7 | 27 ± 4 | 2.1 ± 0.3 |
| 2.6:1 nanorods | 63 ± 8 | 24 ± 3 | 2.6 ± 0.4 |
| 2.9:1 nanorods | 112 ± 10 | 39 ± 3 | 2.9 ± 0.3 |
| 3.4:1 nanorods | 86 ± 7 | 25 ± 2 | 3.4 ± 0.4 |
| 4.0:1 nanorods | 91 ± 5 | 22 ± 1 | 4.0 ± 0.3 |
| 5.6:1 nanorods | 101 ± 8 | 18 ± 2 | 5.6 ± 0.5 |
| 6.8:1 nanorods | 108 ± 14 | 16 ± 2 | 6.8 ± 1.0 |
| nano-bipyramids | 153 ± 3 | 48 ± 2 | 3.2 ± 0.2 |

[a] To calculate the average length-to-diameter aspect ratio $\langle L/d \rangle$, $L/d$ was calculated first for each nanocrystal and then averaged, so in general $\langle L/d \rangle \neq \langle L \rangle / \langle d \rangle$.
Error bars represent the standard deviations of size distributions.

## SUPPLEMENTARY SECTION 2

### MODULATED THERMOREFLECTANCE IMAGING OF LONG-RANGE ANISOTROPY

To visualize anisotropic thermal transport on larger length scales than what is probed in stroboSCAT measurements, we employ frequency domain modulated thermoreflectance at 100 kHz. This experiment was performed on a 15 μm × 20 μm supercrystal domain composed of 4:1 Au–CTAB nanorods. The relative positions of the pump and probe were swept in two dimensions to construct the amplitude and phase profiles below. The major and minor axes were identified and the amplitude and phase profiles were each fit to obtain $D_{\|}/D_{\perp} \approx 4.5$ for this particular supercrystal. This experiment shows that the anisotropic behavior persists out to at least 10 μm despite the likely presence of defects on this length scale.

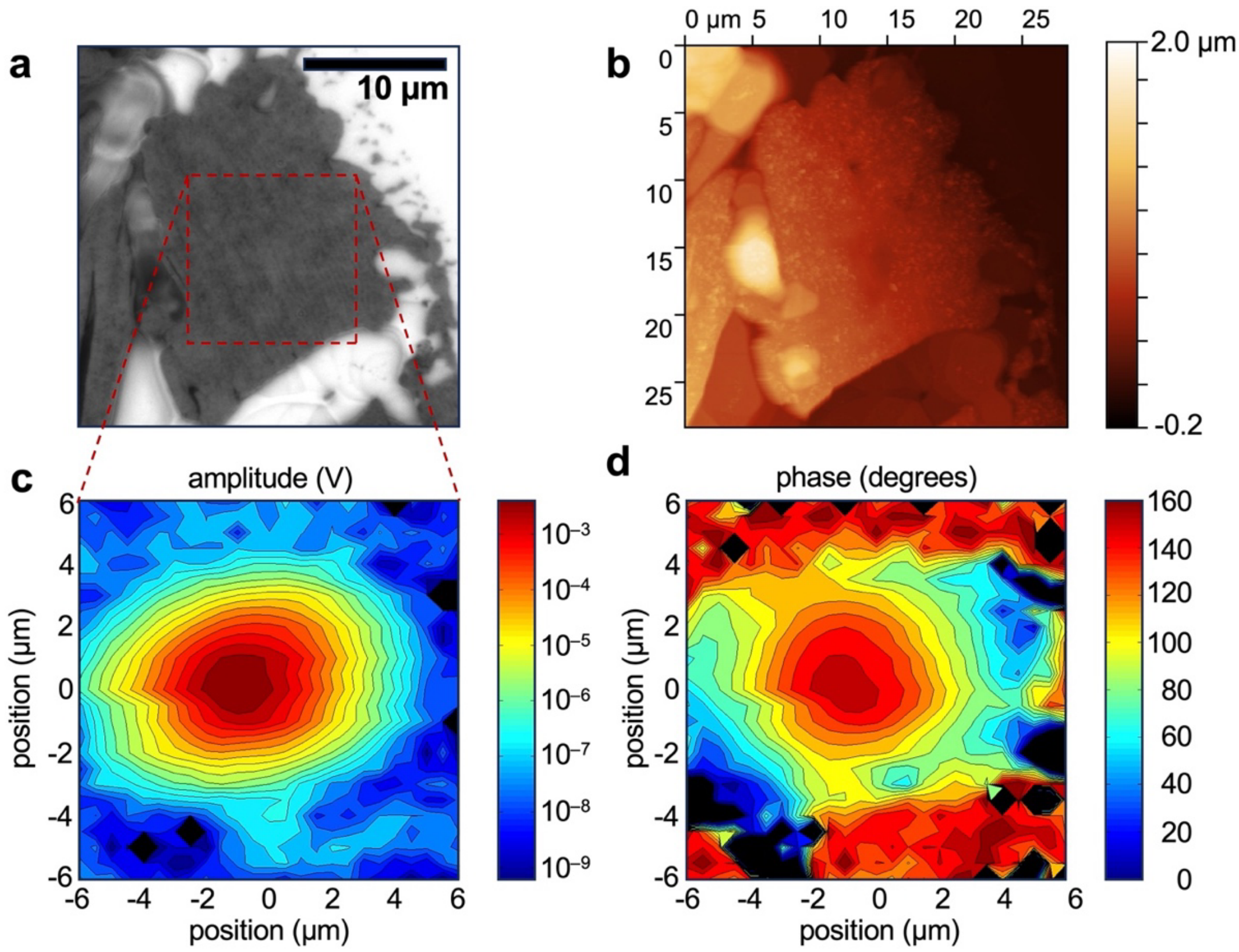

**Figure S4. Modulated thermoreflectance microscopy. a**, Optical reflection image of 4:1 Au–CTAB nanorod supercrystal. **b**, AFM image of the same supercrystal, used for estimating optical absorption length. **c**, Amplitude and **d** phase of modulated thermoreflectance signal for an ordered supercrystal of 4:1 Au–CTAB nanorod supercrystal at 100 kHz, collected at the location indicated in (**a**).

## SUPPLEMENTARY SECTION 3

**CORRELATIVE SEM AND SPATIOTEMPORALLY-RESOLVED THERMOREFLECTANCE**

To achieve correlative SEM and stroboSCAT measurements, we overlaid SEM and optical images and identified distinctive SC features in both to construct a pixel-by-pixel map of the relative coordinates and rotation of the two images (Figure 2). Using a piezoelectric stage on the stroboSCAT optical microscope, we moved to a specific location of choice corresponding to a known location in the SEM image. We then performed stroboSCAT on this location and overlayed $\Delta R/R$ contours on the corresponding SEM images. We estimate that the overlays of the $\Delta R/R$ contours and SEM image are accurate to within about 100–200 nm in lateral displacement and 1° in relative orientation. Elaborating on Figure 2 of the main text, we present SEM images with no overlaid contour and ground-state reflectance images of the corresponding locations.

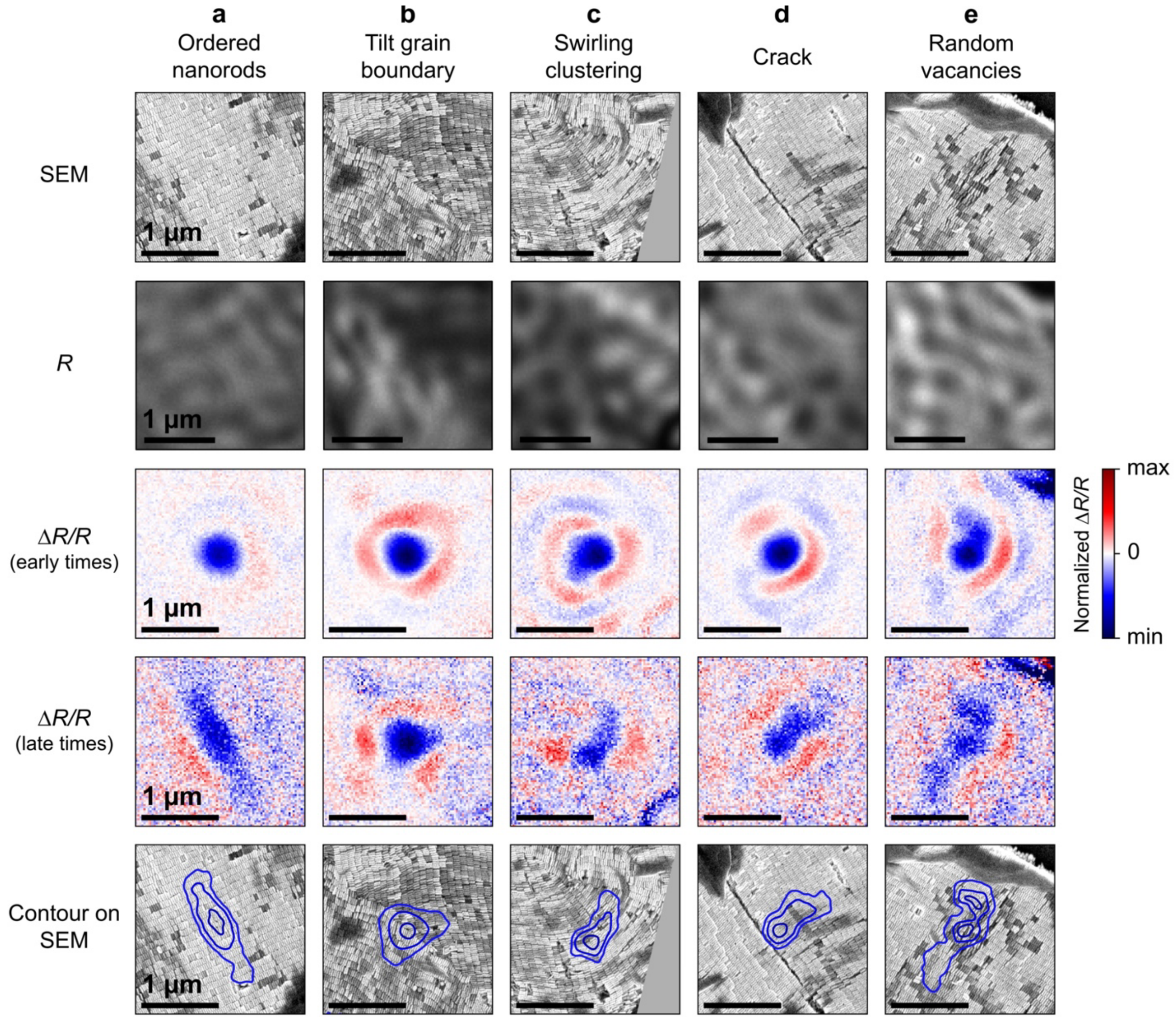

**Figure S5. Supplementary figures for correlative SEM and optical measurements.** Same figures as in Figure 2 for $\Delta R/R$ at early and late times, as well as the 150 ns $\Delta R/R$ contours overlaying the corresponding locations imaged in SEM, but the first two rows show the SEM image without contours overlaid and the ground-state optical reflectance image $R$ collected using a 640 nm probe laser with vertical polarization relative to the images as shown. From left to right, the early time stroboSCAT images correspond to 0.1, 20, 10, 10, and 10 ns after photoexcitation, respectively, and the late-time images are all at 150 ns.

## SUPPLEMENTARY SECTION 4

## SPATIOTEMPORALLY-RESOLVED THERMOREFLECTANCE FOR DIFFERENT NANOROD ASPECT RATIOS

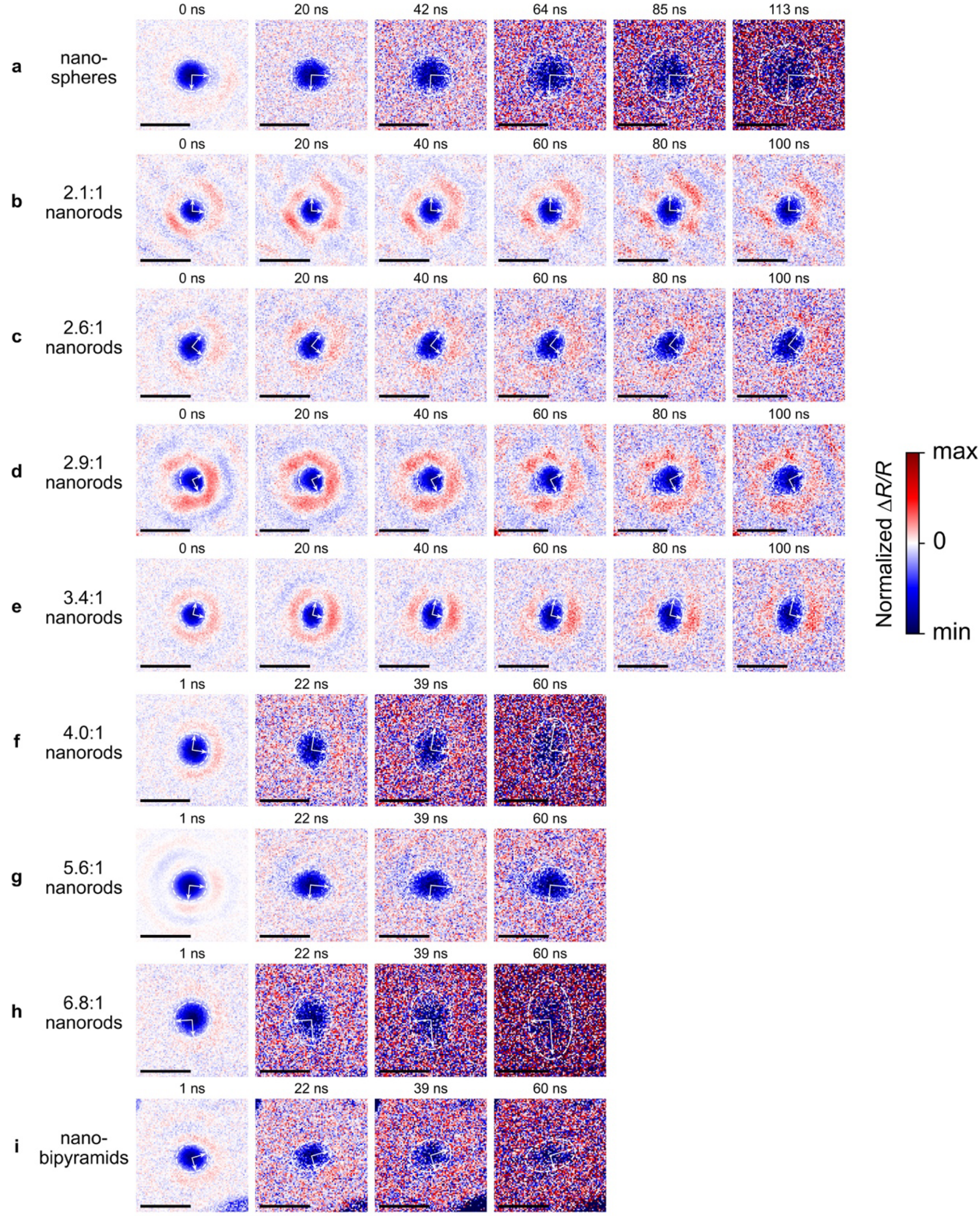

**Figure S6. Representative spatiotemporally-resolved thermoreflectance time series of all samples investigated in this work.** The reported anisotropies reported in the main manuscript (Figure 3) and thermal diffusivities (Table S3) represent the mean and standard deviation of the mean of several such measurements in different locations of the samples.

## SUPPLEMENTARY SECTION 5

## THERMAL DIFFUSIVITY MEASUREMENT RESULTS

**Table S3. Summary of thermal diffusivity measurements**

| Sample Title | $\langle D_\parallel \rangle$ ($\times 10^{-3}$ cm$^2$ s$^{-1}$) | $\langle D_\perp \rangle$ ($\times 10^{-3}$ cm$^2$ s$^{-1}$) | $\langle D_\parallel / D_\perp \rangle$ [a] |
|---|---|---|---|
| nanospheres | 1.4 ± 0.1 | 1.4 ± 0.1 | 1.0 ± 0.1 |
| 2.1:1 nanorods | 0.8 ± 0.1 | 0.41 ± 0.02 | 1.8 ± 0.4 |
| 2.6:1 nanorods | 1.2 ± 0.2 | 0.36 ± 0.05 | 2.7 ± 0.5 |
| 2.9:1 nanorods | 0.9 ± 0.2 | 0.34 ± 0.03 | 2.8 ± 0.3 |
| 3.4:1 nanorods | 1.2 ± 0.2 | 0.38 ± 0.05 | 3.3 ± 0.3 |
| 4.0:1 nanorods | 5.0 ± 0.3 | 1.5 ± 0.2 | 3.2 ± 0.3 |
| 5.6:1 nanorods | 2.7 ± 0.6 | 0.76 ± 0.01 | 4.2 ± 0.3 |
| 6.8:1 nanorods | 9.7 ± 0.5 | 1.7 ± 0.2 | 5.0 ± 0.5 |
| nano-bipyramids | 6.0 ± 0.7 | 1.4 ± 0.2 | 4.4 ± 0.8 |

[a] To calculate $\langle D_\parallel / D_\perp \rangle$, $D_\parallel / D_\perp$ was calculated first for each measurement and then averaged, so in general $\langle D_\parallel / D_\perp \rangle \neq \langle D_\parallel \rangle / \langle D_\perp \rangle$

Error bars represent the standard error of multiple measurements on different locations and different supercrystal domains of each sample.

The moderate variability in $D_\parallel$ and $D_\perp$ yet robust trend in $D_\parallel / D_\perp$ is discussed on page S20.

## SUPPLEMENTARY SECTION 6

### ESTIMATION OF THERMAL CONDUCTIVITIES

The spatiotemporally-resolved thermoreflectance measurement gives a direct measurement of the thermal *diffusivity* along the major and minor axes. To provide values for the corresponding anisotropic thermal conductivities, we calculate $k_{\parallel,\perp} = D_{\parallel,\perp}\rho C_p$ for the major and minor axes using estimated values for the mass density $\rho$ and specific heat capacity $C_p$.[1] Malen and coworkers previously demonstrated[2] that in the case of NC composite solids, the appropriate definitions are the average density—$\rho = \varphi_{NC}\rho_{NC} + (1 - \varphi_{NC})\rho_L$, where $\varphi_{NC}$ is the volume fraction of the NC core and $\rho_{NC}$ and $\rho_L$ are the densities of the NC core and ligand, respectively—and the mass-weighted specific heat capacity—$C_p = m_{NC}C_{NC} + (1 - m_{NC})C_L$, where $m_{NC}$ and $1 - m_{NC}$ are the mass fractions of the NC core and the ligand, respectively, and $C_{NC}$ and $C_L$ are the NC core and ligand specific heat capacities, respectively. In our estimations we assume: $\rho_{Au}$ = 19,300 kg m$^{-3}$, $C_{p,\,Au}$ = 129 J kg$^{-1}$ K$^{-1}$; $\rho_{CTAB/CTAC}$ = 500 kg m$^{-3}$,[3] $C_{p,\,CTAB/CTAC}$ = 1400 J kg$^{-1}$ K$^{-1}$;[4] $\rho_{PSSH}$ = 1000 kg m$^{-3}$,[5] and $C_{p,\,PSSH}$ = 1250 J kg$^{-1}$ K$^{-1}$ based on polystyrene thin films as a proxy.[6] Using these values as common parameters across all samples, we estimate the volume fraction and mass fraction based on the NC dimensions of Table S2. For the nanospheres we use an inter-NC spacing of 4.4 nm based on TEM images, and for all nanorod and bipyramid samples we assume a spacing of 3.4 nm. We assume that excess ligand fills the interstitials for this estimation.

For nanospheres of diameter $d$ and inter-NC spacing $s$ arranged in an FCC structure, the NC volume fraction is given by

$$\varphi_{\text{nanospheres}} = \frac{\frac{4}{3}\pi\left(\frac{d}{2}\right)^3}{\frac{\sqrt{2}}{16}(2d+s)^3}$$

For cylindrical nanorods with semi-spherical tips of length $L$, diameter $d$ and an inter-NC spacing $s$, the NC volume fraction for lateral centered rectangular packing and hexagonal out-of-plane packing writes

$$\varphi_{\text{nanorods}} = \frac{\pi d^2\left(L - \frac{d}{3}\right)}{\sqrt{3}(d+s)^2\left(2L - (2-\sqrt{3})d + \sqrt{3}s\right)}$$

To roughly approximate the volume fraction of Au in the bipyramid SCs, we start with the literature value for the packing fraction of hard pentagonal bipyramids of $\eta$ = 0.835,[7] take this to be the size of the Au particle plus ligand shell, then correct for the volume fraction of Au using the textbook equation for the volume of a perfect pentagonal bipyramid: $\eta V_{NC}/V_{NC+\text{ligand}}$.

The results of these estimations for thermal conductivity appear below. We note that these values should be thought of as approximate due to assumptions made in calculating volume fractions, while the experiment performed here gives a direct measure of the diffusivity.

**Table S4. Summary of thermal conductivities**

| Sample Title | $\rho$ (kg m$^{-3}$) | $C_p$ (J kg$^{-1}$ K$^{-1}$) | $\langle k_{\parallel} \rangle$ (W m$^{-1}$ K$^{-1}$) | $\langle k_{\perp} \rangle$ (W m$^{-1}$ K$^{-1}$) |
|---|---|---|---|---|
| nanospheres | 10,300 | 180 | 0.26 | |
| 2.2:1 nanorods | 11,900 | 150 | 0.14 | 0.073 |
| 2.6:1 nanorods | 12,000 | 150 | 0.22 | 0.065 |
| 2.9:1 nanorods | 13,500 | 143 | 0.18 | 0.066 |
| 3.4:1 nanorods | 12,500 | 147 | 0.22 | 0.070 |
| 4.0:1 nanorods | 12,300 | 148 | 0.91 | 0.27 |
| 5.6:1 nanorods | 11,800 | 151 | 0.48 | 0.14 |
| 6.8:1 nanorods | 11,400 | 152 | 1.7 | 0.30 |
| nano-bipyramids | 14,000 | 140 | 1.2 | 0.28 |

## SUPPLEMENTARY SECTION 7

### FINITE ELEMENT SIMULATION OF THERMAL TRANSPORT ANISOTROPY

The procedure for finite element simulations is described in the Methods. Here, we present representative results for 4:1 Au nanorods as an example. We note that simulations are in two dimensions for computational feasibility, but while the out-of-plane packing can quantitatively change the in-plane anisotropy, our simulations likely capture the essential trends of the system.

Figures S7a and b show example simulation geometries for two different nanorod aspect ratios (4:1 and 2:1, respectively) with side-to-side centered rectangular packing, including interstitial voids due to packing with finite ligand lengths and rounded nanorod tips. The evolution of the initial Gaussian temperature gradient is anisotropic, expanding faster along the axis parallel to the nanorod long-axis (the major axis) than that parallel to the nanorod short-axis (the minor axis) (Figure S7c). Because the thermal conductivity of Au is much larger than that of the organic ligand, the temperature profile is effectively flat throughout the extent of the NC core while temperature drops primarily occur across the ligand gaps (Figure S7d). Mirroring the experiments, we extract the temperature profiles along each axis (Figure S7d), fit to a Gaussian profile over time to obtain the mean squared displacement (Figure S7e) and retrieve the diffusivity along each axis.

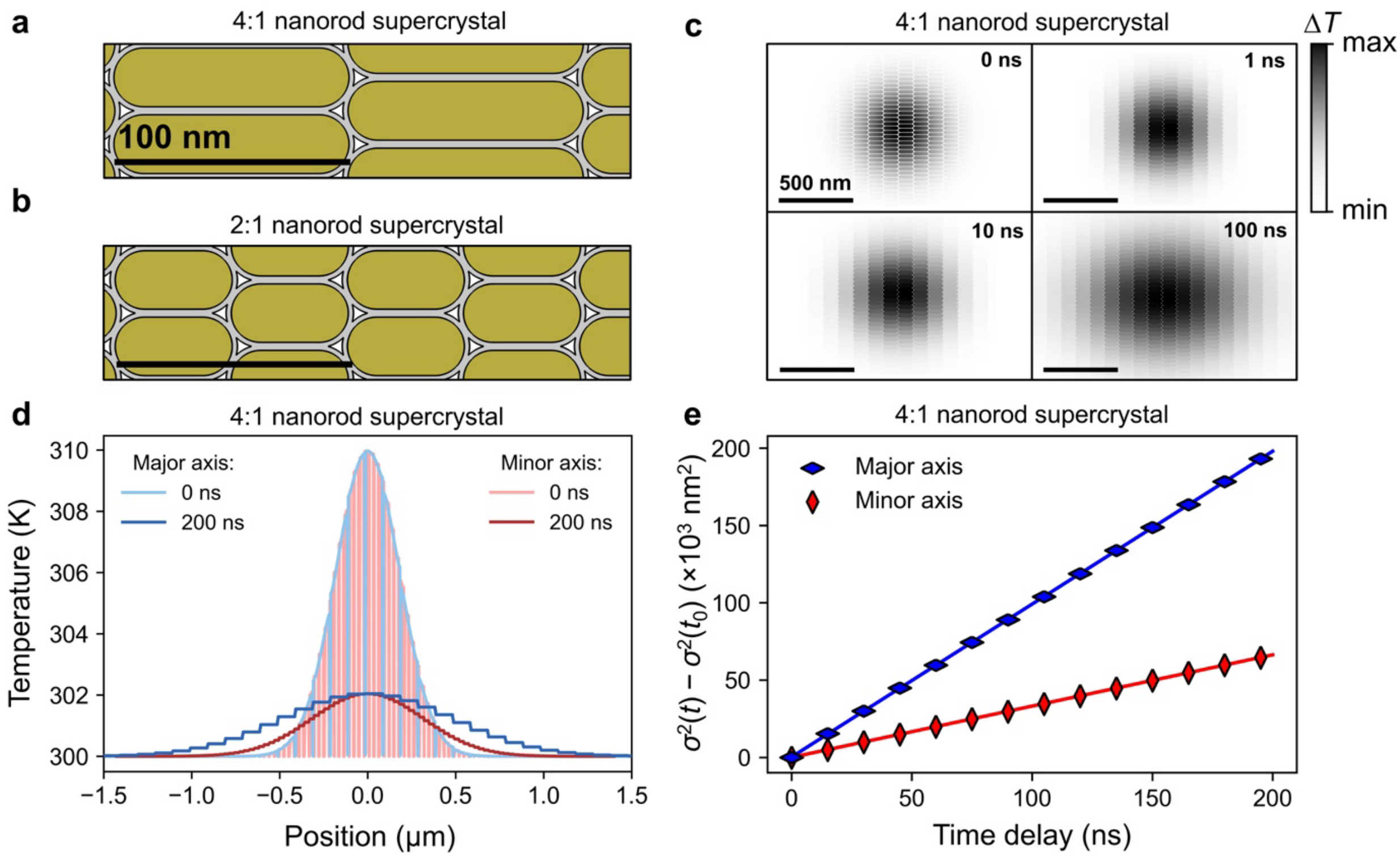

**Figure S7. Finite element simulation of thermal transport. a**, Simulation geometry, featuring 100 nm × 25 nm (4:1) nanorods with rounded tips, arranged in a side-by-side, centered rectangular packing structure as seen in SEM images. Gold color indicates the Au, gray indicates the CTAB ligand, and white indicates air voids. **b**, Simulation geometry for 50 nm × 25 nm (2:1) nanorods with rounded tips. **c**, Representative simulated temperature map evolution over time for 4:1 nanorods, starting from a 420 nm (fwhm) Gaussian profile in the Au nanorods cores. **d**, Temperature profiles along the major and minor axis line cuts at 0 and 200 ns for 4:1 nanorods. Initially the temperature rise is restricted to the Au NC cores while the ligands remain at ambient temperature, but after time the radial temperature profiles are monotonic with flat profiles within the NCs and temperature drops across the ligand barriers. **e**, Representative simulated mean-squared expansion curves along the major and minor axes for 4:1 nanorods.

Results of the finite element simulations are summarized in Table S5.

**Table S5. Summary of simulation results**

| Sample Title | With voids | | | Without voids | | |
|---|---|---|---|---|---|---|
| | $D_{\parallel}$ ($\times 10^{-3}$ cm$^2$ s$^{-1}$) | $D_{\perp}$ ($\times 10^{-3}$ cm$^2$ s$^{-1}$) | $D_{\parallel}/D_{\perp}$ | $D_{\parallel}$ ($\times 10^{-3}$ cm$^2$ s$^{-1}$) | $D_{\perp}$ ($\times 10^{-3}$ cm$^2$ s$^{-1}$) | $D_{\parallel}/D_{\perp}$ |
| nano-spheres (41 nm) | – | – | – | 2.5 | 2.5 | 1 |
| 2:1 nanorods (50 nm × 25 nm) | 2.6 | 1.6 | 1.6 | 3.2 | 1.7 | 1.9 |
| 3:1 nanorods (75 nm × 25 nm) | 3.8 | 1.6 | 2.3 | 4.7 | 1.7 | 2.7 |
| 4:1 nanorods (100 nm × 25 nm) | 4.9 | 1.7 | 3.0 | 6.1 | 1.7 | 3.6 |
| 5:1 nanorods (125 nm × 25 nm) | 6.1 | 1.7 | 3.7 | 7.6 | 1.7 | 4.4 |
| 6:1 nanorods (150 nm × 25 nm) | 7.3 | 1.7 | 4.4 | 9.1 | 1.7 | 5.3 |
| 7:1 nanorods (175 nm × 25 nm) | 8.5 | 1.7 | 5.0 | 10.6 | 1.7 | 6.1 |
| 3:1 nano-bipyramids (153 nm × 48 nm) | 12.2 | 1.9 | 6.5 | 13.8 | 2.0 | 7.0 |

Simulation geometries for all nanorod samples are the side-to-side packing geometry in Figure 3f.
Simulation geometries for all nanorod and nano-bipyramid samples used thermal parameters of CTAB ligands (Methods) with a closest inter-NC spacing of 3.4 nm and ligand length of 2.2 nm.
Simulation geometries for nanospheres used thermal parameters of PSSH ligands (Methods) with an inter-NC spacing of 4.4 nm based on TEM images.

## SUPPLEMENTARY SECTION 8

## EFFECTIVE MEDIUM APPROXIMATION MODELING OF SERIES RESISTANCE

The success of the finite element simulations here (Figure 3a) and previous[2,8–11] effective medium modeling of isotropic NC films motivates developing an analytical equivalent thermal circuit model for the thermal transport anisotropy. The finite element simulations are the most general modeling approach in this work, but they do not provide a means to easily identify trends and estimate how thermal transport anisotropy depends on the thermal and morphological parameters of the system. Here we derive a simple analytical approximation for the thermal transport anisotropy and use it to examine the behavior and limits of the phenomenon. While this model does not capture the influence of NC packing geometry, which also plays an important role in the final anisotropy as we discuss in the main text and the following sections, it provides a convenient toy model to gain insights into general behavior of thermal anisotropy in NC solids.

### Effective thermal conductivity of a flat plate composite

Consider the one-dimensional heat transport problem of a plate of thickness $a$ and area $A$ perpendicular to heat flow (Figure S8a).

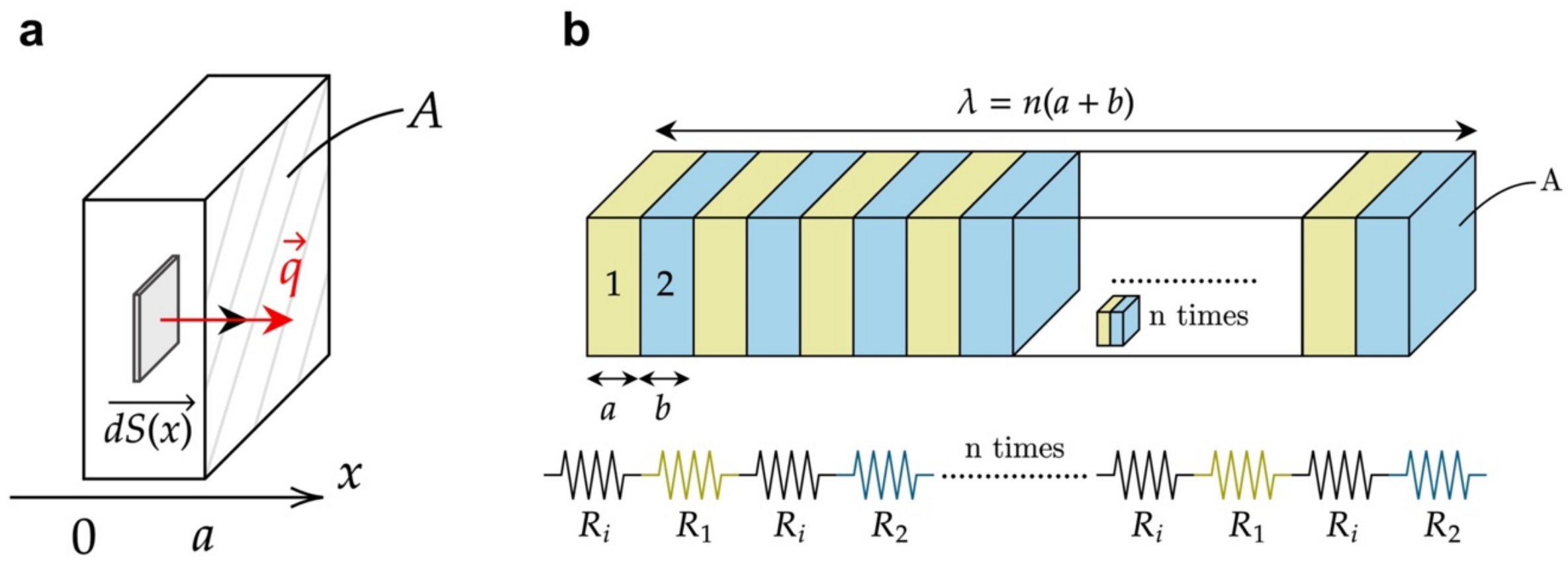

**Figure S8. Diagram of flat plates and heat flow. a,** Flat plate of thickness $a$ and area $A$ with heat flux and surface element d$S$. **b,** Flat plate composite medium with corresponding equivalent thermal circuit.

The isothermal surfaces are planes, the heat flux density $q$ is parallel to the $x$-direction, and the temperature gradient $\Delta T$ depends only on $x$. We calculate the temperature drop across the plate using Fourier law:

$$|\Delta T| = \int_0^a \frac{q}{k}\mathrm{d}x = \frac{qa}{k}$$

where $k$ is the thermal conductivity of the material. The total heat flux going through the wall is:

$$Q = \iint_A \vec{q} \cdot \mathrm{d}\vec{S} = qA$$

Thus, the thermal resistance of the plate writes:

$$R = \frac{|\Delta T|}{Q} = \frac{a}{kA} \quad \text{(S1)}$$

Now, consider a composite medium made of alternating flat plates of materials 1 and 2 (Figure S8b). Because of the symmetry between Ohm's law and Fourier law, we can treat the composite material as a thermal equivalent circuit. In this case, we have a circuit with series resistances due to the two materials and the interfaces between them. The total resistance writes:

$$R_{\text{tot}} = nR_1 + nR_2 + 2nR_i \quad \text{(S2)}$$

Where $R_1$ and $R_2$ are the thermal resistances of plates 1 and 2, $R_i$ is the thermal interfacial resistance between the two plates, and $n$ is the number of unit cells made of plates 1 and 2. Substituting the expression of the flat plate resistance from equation (S1), equation (S2) becomes:

$$\frac{\lambda}{k_{\text{eff}}A} = n\frac{a}{k_1 A} + n\frac{b}{k_2 A} + 2n\frac{1}{h_c A} \quad \text{(S3)}$$

Where $k_{\text{eff}}$ is the effective thermal conductivity of the composite, $k_1$ and $k_2$ are the thermal conductivities of materials 1 and 2, and $h_c$ is the contact thermal conductance per unit area.

We now define the material 1 and 2 volume fractions $\phi_1$, $\phi_2$ in the composite as:

$$\phi_1 = \frac{nAa}{A\lambda} = \frac{na}{\lambda} = \frac{a}{a+b}$$

$$\phi_2 = \frac{nAb}{A\lambda} = \frac{\lambda - na}{\lambda} = 1 - \phi_1$$

We can rewrite equation (S3) in terms of material volume fractions:

$$\frac{1}{k_{\text{eff}}} = \frac{\phi_1}{k_1} + \frac{1-\phi_1}{k_2} + \frac{2\phi_1}{ah_c} \quad \text{(S4)}$$

This expression is the result obtained by Hasselman and Johnson in their work about effective thermal conductivity of composites.[11]

**One-dimensional model of a nanorod supercrystal**

Consider a supercrystal of rectangular nanorods arranged in a grid-like pattern as in Figure S9. In order to analytically model the thermal anisotropy of this composite, we apply the above one-dimensional flat plate composite model to each one of the major (parallel to the nanorod long axis, $\parallel$) and minor (perpendicular to the nanorod long axis, $\perp$) axes. In this model, the real three-dimensional geometry is approximated as two perpendicular one-dimensional models to obtain the thermal diffusivity ratio. This assumes that thermal transport along these two directions is independent and does not account for diagonal thermal paths between the nanorods that are possible in a higher dimensional geometry.

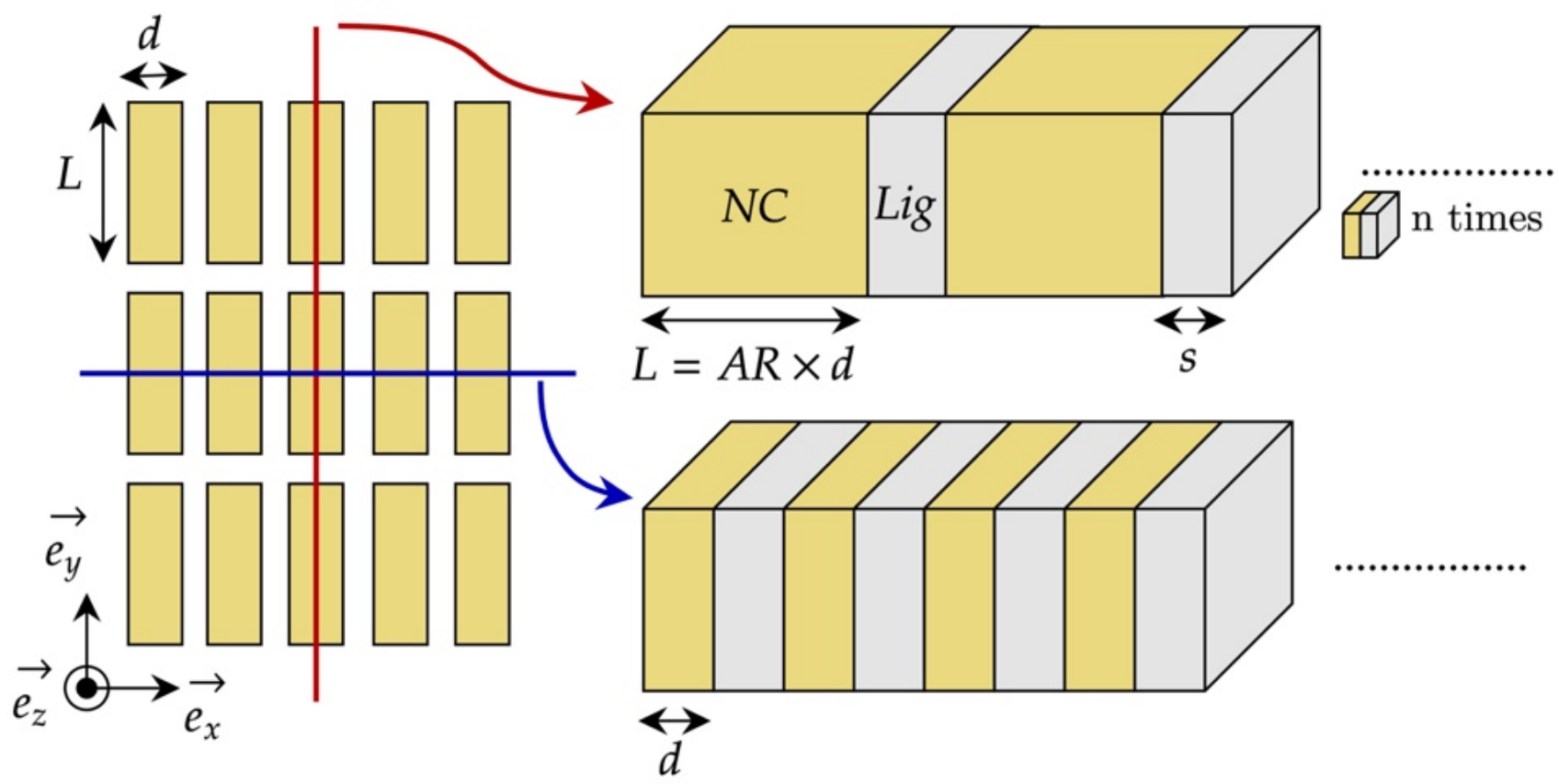

**Figure S9. Diagram of a supercrystal with grid-like packing of rectangular nanorods.** The major and minor axes (blue and red lines, respectively) are modeled by two different flat plate composites with different NC thicknesses $L$ and $d$, giving an effective aspect ratio of AR = $L/d$. Regions with the thermal properties of the NC core material and the ligand material are shown as gold and gray, respectively.

The fact that the NC core material volume fraction is now different along the major and minor axes will lead to different thermal conductivities along these directions. The NC core material volume fractions write:

$$\phi_{\parallel} = \frac{L}{L+s} \qquad \phi_{\perp} = \frac{d}{d+s}$$

where $L$ and $d$ are the "length" (long dimension) and the "diameter" (short dimension) of the nanorod and $s$ is the spacing between adjacent nanorods that is filled with uniform ligands.

Using equation (S4) we obtain the thermal anisotropy as:

$$\frac{D_{\parallel}}{D_{\perp}} = \frac{k_{\parallel}}{k_{\perp}} = \frac{\frac{\phi_{\perp}}{k_{\mathrm{NC}}} + \frac{1-\phi_{\perp}}{k_{\mathrm{L}}} + \frac{2\phi_{\perp}}{dh_c}}{\frac{\phi_{\parallel}}{k_{\mathrm{NC}}} + \frac{1-\phi_{\parallel}}{k_{\mathrm{L}}} + \frac{2\phi_{\parallel}}{Lh_c}} = \frac{2\frac{k_{\mathrm{NC}}}{dh_c} + \frac{k_{\mathrm{NC}}}{k_{\mathrm{L}}}\frac{s}{d} + 1}{2\frac{k_{\mathrm{NC}}}{dh_c} + \frac{k_{\mathrm{NC}}}{k_{\mathrm{L}}}\frac{s}{d} + \frac{L}{d}} \cdot \frac{\frac{s}{d} + \frac{L}{d}}{\frac{s}{d} + 1} \tag{S5}$$

According to this equation, the anisotropy depends on four dimensionless parameters: the length-to-diameter aspect ratio $L/d$, the spacing-to-diameter ratio $s/d$, the thermal conductivity contrast between the NC and ligand materials $k_{\mathrm{NC}}/k_{\mathrm{L}}$, and the thermal conductivity contrast between the NC material and the NC–ligand interface $k_{\mathrm{NC}}/dh_{\mathrm{c}}$.

Finite element simulations were performed to verify this model's validity. We simulated the ratio of two one-dimensional heat transport problems to compare to equation (S5) (two NC composites with flat plates of thicknesses $L$ and $d$), a two-dimensional supercrystal of rectangular nanorods arranged in a grid, a two-dimensional supercrystal of rounded nanorods in a side-by-side centered rectangular pattern in a ligand matrix (geometry of Figure 3f) and the same arrangement with interstitial voids (geometry of Figure 3f). Simulations that adhere to the geometry of the one-dimensional model perfectly agree with it (Figure S10), confirming that equation (S5) is valid within its assumptions. Adding an extra dimension to the problem by performing a finite element simulation of a grid of rectangles slightly increases the anisotropy with respect to the plate model because diagonal paths that mainly contribute to major axis transport become available. Going from rectangles to rounded nanorod tips decreases the anisotropy because the effective NC volume fraction along the major axis decreases. Using finite length ligand shells further reduces the anisotropy due to the nonconductive voids that form near the nanorod tips. Overall, these simulations indicate how the simplistic analytical model breaks down when more realistic morphological details are included; yet, the analytical model appears to reproduce the essential behavior of thermal anisotropy in such supercrystals, motivating its use to explore trends.

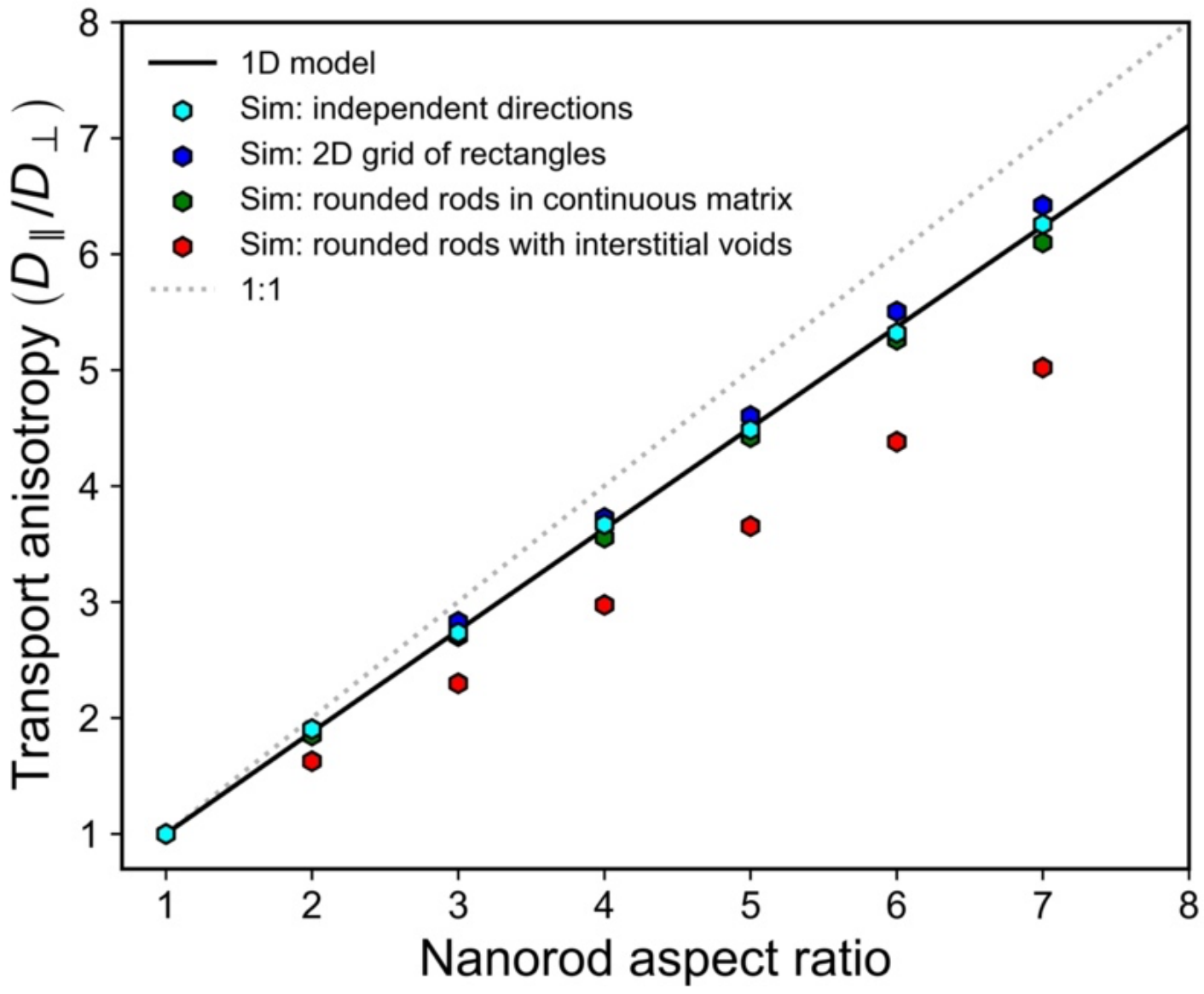

**Figure S10. Comparison of thermal transport anisotropy between the series resistance model and simulations.** In the case of the "independent directions" simulation, we constructed two separate flat plate composites with NC thicknesses $L$ and $d$ and divided the resulting diffusivities whereas we simulated the complete supercrystal in the other simulations. All simulations have been performed with the same thermal and morphological parameters as the Au–CTAB nanorod samples studied here: $k_{\mathrm{NC}} = 310$ W m$^{-1}$ K$^{-1}$, $k_{\mathrm{L}} = 0.15$ W m$^{-1}$ K$^{-1}$, $h_{\mathrm{c}} = 40$ MW m$^{-2}$ K$^{-1}$, $s = 3.4$ nm, $d = 25$ nm.

**Influence of parameters and emergent trends**

Here we examine the behavior of the series resistance model as a function of the relevant parameters.

*Aspect ratio* ($L/d$). According to this model, the thermal transport anisotropy $D_{\parallel}/D_{\perp}$ monotonically increases with the aspect ratio $L/d$ (Figure S11) but has a saturation limit imposed by the relative sizes of the ligand gap and NC of the supercrystal and the relative thermal properties of the NC and the ligands. In the limit of large aspect ratio, the anisotropy saturates at a value of:

$$\lim_{L/d\to\infty}\left(\frac{D_{\parallel}}{D_{\perp}}\right) = \frac{2\cdot\frac{k_{\mathrm{NC}}}{dh_c} + \frac{k_{\mathrm{NC}}}{k_{\mathrm{L}}}\cdot\frac{s}{d} + 1}{\frac{s}{d} + 1}$$

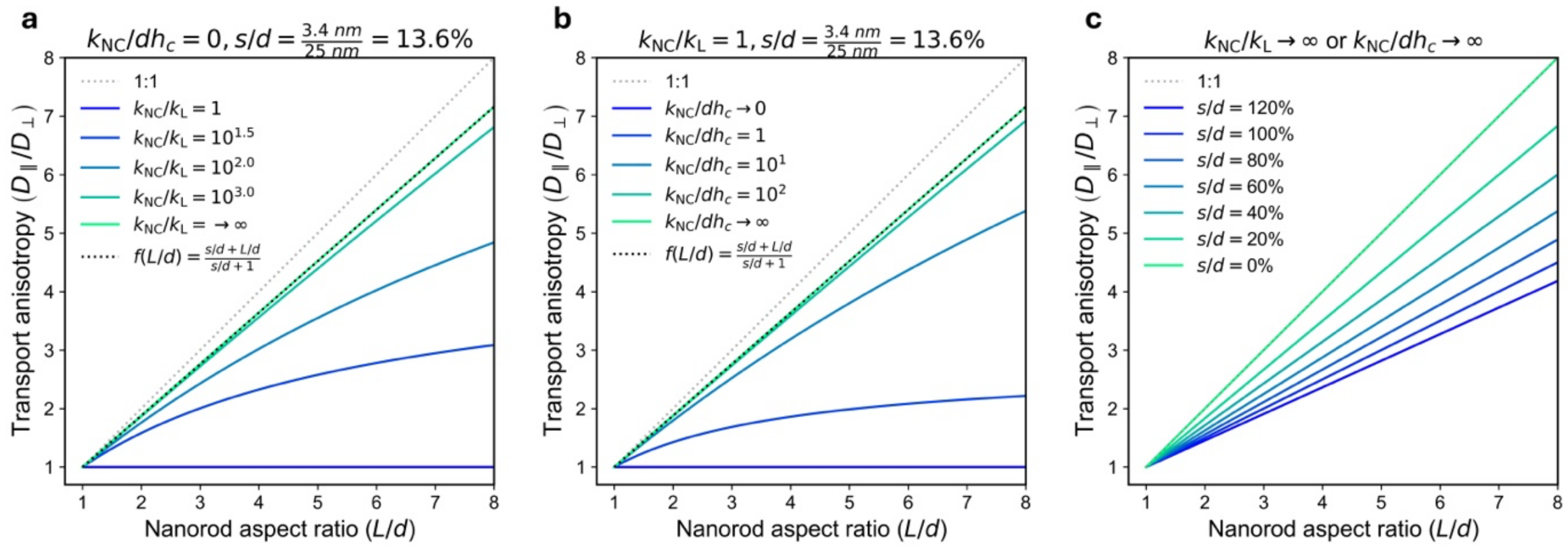

**Figure S11. Limits of transport anisotropy as a function of aspect ratio for different parameter values. a,** Anisotropy limit in absence of interfacial resistance as a function of aspect ratio ($dh_c \gg k_{NC}$) for values of $k_{NC}/k_L$ ranging from 1 to $10^3$. **b,** Anisotropy limit in absence of conductivity contrast between the NC material and the ligands ($k_{NC} = k_L$) for values of conductivity contrast between the NC material and the NC-ligand interface ranging from 0 to $10^2$. **c,** Anisotropy limit with maximum conductivity contrast between the NC material and the ligands or the ligand–NC interface for $s/d$ ratios ranging from 0% to 120%.

*Material conductivities* ($k_{NC}$, $k_L$) *and interfacial thermal conductance* ($h_c$). In the absence of an interfacial thermal resistance between the nanorod material and the ligands, in other words when the interfacial conductivity is much greater than the NC conductivity ($k_{NC}/dh_c \rightarrow 0$), the thermal anisotropy reduces to:

$$\frac{D_\parallel}{D_\perp} = \frac{\frac{k_{NC}}{k_L} \cdot \frac{s}{d} + 1}{\frac{k_{NC}}{k_L} \cdot \frac{s}{d} + \frac{L}{d}} \cdot \frac{\frac{s}{d} + \frac{L}{d}}{\frac{s}{d} + 1}$$

Increasing the conductivity contrast between the NCs and the ligands from $k_{NC}/k_L = 1$ to $\infty$ will drive $D_\parallel/D_\perp$ from 1 to an upper limit of:

$$\frac{D_\parallel}{D_\perp} = \frac{\frac{s}{d} + \frac{L}{d}}{\frac{s}{d} + 1} = \frac{\phi_\perp}{\phi_\parallel} \cdot \frac{L}{d} \quad \text{(S6)}$$

In other words, when the contrast between the NC and ligand conductivities is sufficiently high, the anisotropy becomes limited by the composite's geometry (the limit depends only on the geometric parameters $s$, $d$, and $L$). This limit tends to the nanorod aspect ratio itself when the NC spacing $s$ becomes negligible with respect to the NC characteristic size $d$ (Figure S11a).

$$\lim_{s/d \to 0} \left( \lim_{k_{NC}/k_L \to \infty} \left( \frac{D_\parallel}{D_\perp} \right) \right) = \frac{L}{d}$$

It is important to remember that this limit only applies to grid-like packing of NCs and can be exceeded in other packing geometries not captured by this one-dimensional model. Such cases are described in the following sections.

Similarly, in the absence of conductivity contrast ($k_{NC}/k_L = 1$), the thermal anisotropy reduces to:

$$\frac{D_{\parallel}}{D_{\perp}} = \frac{2 \cdot \frac{k_{\mathrm{NC}}}{dh_c} + \frac{s}{d} + 1}{2 \cdot \frac{k_{\mathrm{NC}}}{dh_c} + \frac{s}{d} + \frac{L}{d}} \cdot \frac{\frac{s}{d} + \frac{L}{d}}{\frac{s}{d} + 1}$$

Increasing the contrast between the NC material conductivity and the NC–ligand interfacial conductivity from 0 to ∞, drives the anisotropy from 1 to an upper limit of

$$\frac{D_{\parallel}}{D_{\perp}} = \frac{\frac{s}{d} + \frac{L}{d}}{\frac{s}{d} + 1} = \frac{\phi_{\perp}}{\phi_{\parallel}} \cdot \frac{L}{d}$$

as for large $k_{\mathrm{NC}}/k_{\mathrm{L}}$ (Figure S11b). The anisotropy is thus limited by the geometry in the same manner whether the relative interfacial resistance is high or the contrast between the NC and ligand conductivities is high.

*Inter-NC spacing* (*s*) *and NC size* (*d*). The anisotropy limit reached when $k_{\mathrm{NC}}/k_{\mathrm{L}} \rightarrow \infty$ or $k_{\mathrm{NC}}/dh_{\mathrm{c}} \rightarrow \infty$ is governed by the geometric ratio $s/d$, increasing as $s/d$ decreases (Figure S11c). In other words, for a given ligand, supercrystals that feature short ligands or large nanorods have the potential to exhibit a higher anisotropy than supercrystals of smaller nanorods.

*Relationships between parameters.* Here we examine the anisotropy attainable as a function of selected parameters. Figure S12 shows the anisotropy (for different NC sizes and aspect ratios) as a function of the thermal conductivity of the nanorod core material and the interfacial thermal conductance between the nanorods and the ligands. For these plots we take $k_{\mathrm{L}} = 0.15$ W m$^{-1}$ K$^{-1}$ to be representative of hydrocarbon ligands, a ligand spacing of $s = 3$ nm as an example relevant to CTAB or oleic acid ligands, $h_{\mathrm{c}}$ ranging from 40 to 400 MW m$^{-2}$ K$^{-1}$ to span the range of hydrocarbon–metallic or semiconductor interfaces,[2,8,10,12] and $k_{\mathrm{NC}}$ ranges over 1–400 W m$^{-1}$ K$^{-1}$ to span the range of typical NC core materials. The dashed contour lines highlight the values of $h_{\mathrm{c}}$ and $k_{\mathrm{NC}}$ for which the anisotropy reaches a given percentage of its maximal geometric limit of equation (S6). The color gradient hardly changes along the $h_{\mathrm{c}}$ axis showing that in the range of possible $h_{\mathrm{c}}$ values, the anisotropy is dominated by the thermal conductivity of the nanorod material rather than by the thermal interfacial conductance. For smaller nanorods with $d = 5$ nm (diameter of the same order as NC spacing), we can achieve an anisotropy close to 90% of the geometric limit for aspect ratios $L/d \leq 10$ with a nanorod thermal conductivity of the order of $k_{\mathrm{NC}} \approx 10$ W m$^{-1}$ K$^{-1}$. For larger nanorods (*d* ten times larger), we rather need a NC thermal conductivity of the order of $k_{\mathrm{NC}} \approx 100$ W m$^{-1}$ K$^{-1}$ to reach 90% of maximal anisotropy for aspect ratios ≤ 10.

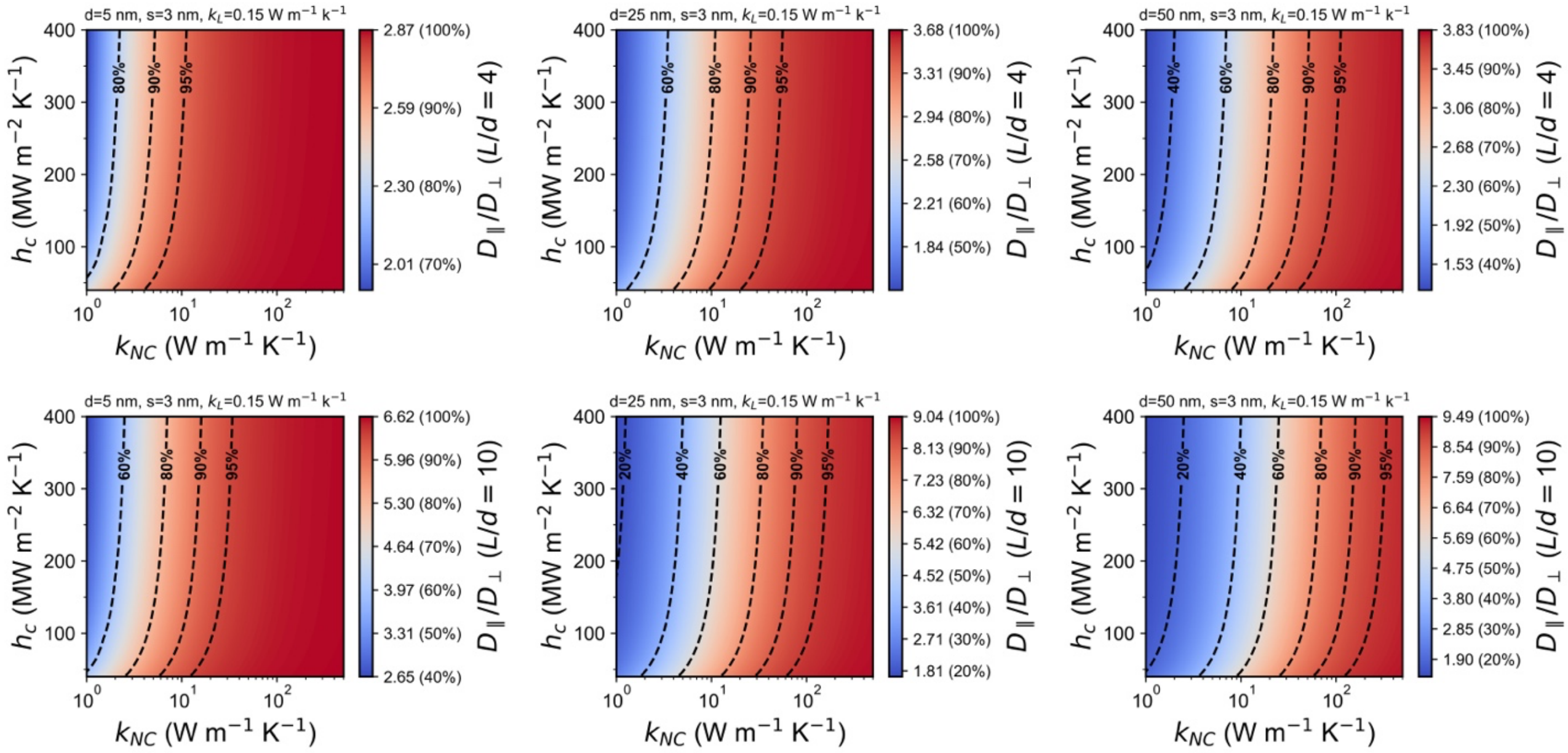

**Figure S12. Thermal transport anisotropy as a function of nanocrystal thermal conductivity and interfacial conductance per unit area for different sizes and aspect ratios.** The thermal transport anisotropy is calculated using equation (S5) as a function of NC core thermal conductivity and interfacial conductance assuming the same ligand conductivity and inter-NC spacing as the Au–CTAB nanorod samples studied here, an aspect ratio of $L/d = 4$ and $L/d = 10$ as well as diameters of 5, 25 and 50 nm.

## Expected anisotropy in common nanocrystal systems

As mentioned in the main text, we expect that the thermal anisotropy we observe in Au NC supercrystals should persist in most common NC core materials. Specifically, using the series resistance model we find that—for the same ligand conductivity and interfacial resistance of the Au–CTAB system—the anisotropy would decrease by only about ~20% when using a NC core material with thermal conductivity even 100 times smaller than gold (Figure S13a), which encompasses many semiconductors, metal oxides, and metals. SCs made of smaller nanorods have a lower maximum anisotropy limit (equation (S5), Figure S12), but values close to the said limit are attained for a larger range of thermal properties; in SCs made of smaller nanorods (e.g., $d \approx 5$ nm relevant to CdS and PbS nanorods) a $k_{NC}$ as low as 10 W $m^{-1}$ $K^{-1}$ suffices to reach 90% of the maximum anisotropy for $L/d \leq 10$ while a NC conductivity of the order of 100 W $m^{-1}$ $K^{-1}$ is required for SCs of wider nanorods ($d \approx 25$ nm) (Figure S12).

Commonly used hydrocarbon ligands in NC systems typically induce inter-NC spacings in the range of 0.5 to 5 nm. The series resistance model predicts a relatively small change in the anisotropy for our Au nanorods over this ligand length range (Figure S13b), increasing for shorter ligands. The dependence is stronger for smaller diameter nanocrystals. In the hypothetical limit of removing the ligand entirely, the thermal anisotropy would still approach the constituent aspect ratio if a substantial interfacial resistance were maintained in the form of grain boundaries between the original NC cores; this could be an interesting approach to achieve tailored thermally anisotropic materials but with high thermal conductivity.

Overall, this estimation is promising for the potential to achieve significant thermal anisotropy for a wide range of anisotropic NC solids.

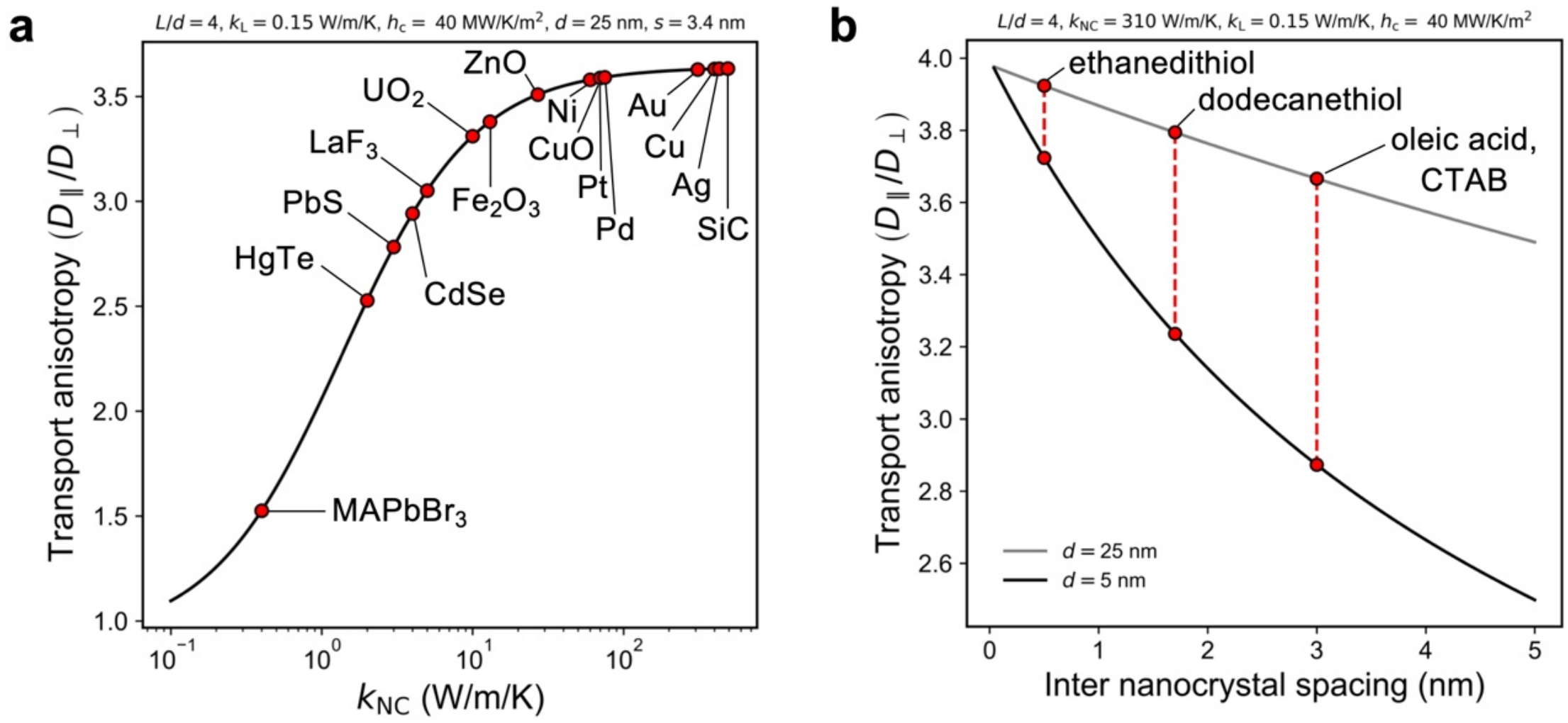

**Figure S13**. **Calculated thermal transport anisotropy for common nanocrystal core materials and ligand lengths. a**, Thermal transport anisotropy calculated using equation (S5) as a function of NC core thermal conductivity assuming an aspect ratio of $L/d$ = 4 and the same thermal and morphological parameters as the Au–CTAB nanorod samples studied here with $k_L$ = 0.15 W m$^{-1}$ K$^{-1}$, $h_c$ = 40 MW m$^{-2}$ K$^{-1}$, $d$ = 25 nm and $s$ = 3.4 nm. The bulk thermal conductivities of several common NC materials are indicated for comparison. **b**, Thermal transport anisotropy calculated using equation (S5) as a function of inter-NC spacing assuming an aspect ratio of $L/d$ = 4 and the same thermal parameters as the Au–CTAB nanorod samples studied here with $k_{NC}$ = 310 W m$^{-1}$ K$^{-1}$ and $k_L$ = 0.15 W m$^{-1}$ K$^{-1}$, $h_c$ = 40 MW m$^{-2}$ K$^{-1}$. This value of $k_L$ is similar to other organic materials. We compare $d$ = 5 and 25 nm. The typical inter-NC spacing produced by common ligands are indicated for comparison.

**Summary of emergent trends and design principles based on series resistance model**

This one-dimensional model allows us to understand the role of the system's physical parameters on the thermal anisotropy. First of all, the anisotropy monotonically increases with aspect ratio before eventually reaching a maximum value. For a given aspect ratio, the thermal anisotropy is limited by the geometry and increases as the NC spacing becomes negligible with respect to the NC diameter. Thus, generally speaking, the smaller the inter-NC spacing or the larger the NC the higher the maximum achievable thermal anisotropy. Moreover, in order to achieve maximum anisotropy within a given supercrystal geometry, one could maximize the NC core thermal conductivity with respect to either that of the ligand or the interfacial thermal conductivity. However, in the case of colloidal nanorod SCs with typical organic ligands, the conductivity contrast between the NC core and the ligands dominates the thermal anisotropy because common interfacial conductances are already relatively large. Although supercrystals made of small nanorods have a lower maximum anisotropy limit, values close to the said limit are easier to reach. It is also interesting to note that moderate variations in factors such as ligand density (that would in turn change $k_L$ and $h_c$) and SC packing density (that would affect the ratio $s/d$) can cause significant changes to the magnitudes of $D_{\parallel}$ and $D_{\perp}$ but have only a minor effect on the resulting anisotropy (Figure S11, S12, S13); this can help rationalize the considerable sample-to-sample variability seen in the experimentally measured diffusivities that nevertheless exhibit a robust, monotonic trend in $D_{\parallel}/D_{\perp}$ (Table S3).

As discussed in the main text and below, the NC shape and packing arrangement also controls anisotropy in ways not captured by this analytical model, but we expect that while the specific limits such as $D_{\parallel}/D_{\perp} \leq L/d$ to not be globally true, the above general trends to increase anisotropy to hold.

## SUPPLEMENTARY SECTION 9

## INCREASED THERMAL TRANSPORT ANISOTROPY THROUGH NANOCRYSTAL SHAPE AND PACKING

### Simulations of nano-bipyramid supercrystals

We simulated 3.2:1 bipyramid-shaped NC SCs by taking a two-dimensional cross section as a model to explore the impact of this NC shape on thermal transport anisotropy (Figure S14). For a continuous ligand matrix, the anisotropy reaches 7.0. Replacing the continuous ligand matrix by well-defined ligand shell of 2.2 nm with 3.4 nm closest inter-NC spacing leads to a marginal decrease of the anisotropy down to 6.6; a finite ligand length induces non-conductive voids between the bipyramid tips that primarily hinder transport along the major axis direction for this packing structure. We note that the two-dimensional geometry simulated here does not reflect the complexity of packing in three dimensional pentagonal bipyramids so we do not expect quantitative agreement with the experiments in Figure 3,[7] but the geometry captures the essential behavior of this NC shape—correctly predicting an anisotropy exceeding the NC aspect ratio.

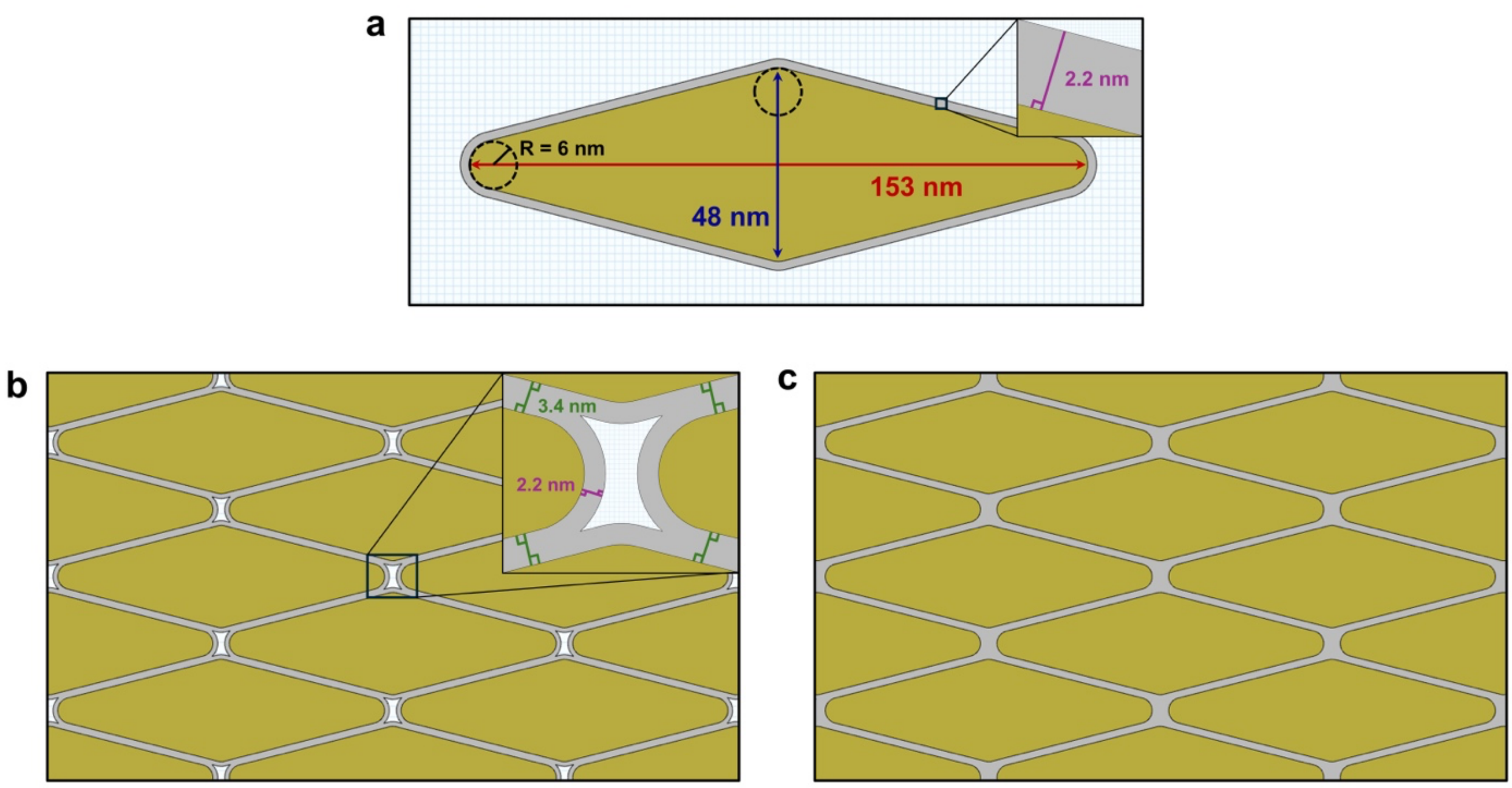

**Figure S14**. **Enhanced thermal transport anisotropy in nano-bipyramid supercrystals.** Simulation geometry approximating two-dimensional cross section of (**a**) nano-bipyramid supercrystals both allowing for (**b**) interstitial voids due to finite ligand lengths and for (**c**) continuous, filled interstices.

### Nanorod supercrystal packing with anisotropy exceeding aspect ratio

The densely-packed, side-by-side "smectic-like" centered rectangular nanorod supercrystal structure primarily featured in the main text appears to be the most common in our samples (Figure 1c, 2c, 3c, S15a,c) and in past reports. However, other nanorod packing structures are possible. We explore the case of end-to-end packing with staggered adjacent rows, resembling a brick wall pattern, which is more rare yet appears in some regions of our SCs (Figure S15b,d) and has been reported previously.[13] We take the case of 4:1 nanorods with a diameter of 25 nm as an example; while the side-by-side packing geometry (Figure S15a) induces a thermal anisotropy of 3.6 (lower than aspect ratio), the brick wall geometry (Figure S15b) induces a thermal anisotropy of 5.5 (40% higher than the nanorod aspect ratio). We attribute this behavior to coupling between the two dimensions of space: new efficient thermal pathways can turn small advances along the slow-axis into additional contributions along the fast-axis that bypass major-axis ligand barriers (Figure S16a,b). In fact, this change in anisotropy is driven by an increase in the major-axis thermal diffusivity while the minor axis thermal diffusivity stays the same. A physical picture for this behavior is described in the section below.

We note that the nanorod arrangement in Figure S15b,d is symmetry related to the nano-bipyramid supercrystal structure in Figure S14. The difference in their resulting thermal transport anisotropies suggests that not only NC arrangement, but also the NC shape plays an important role.

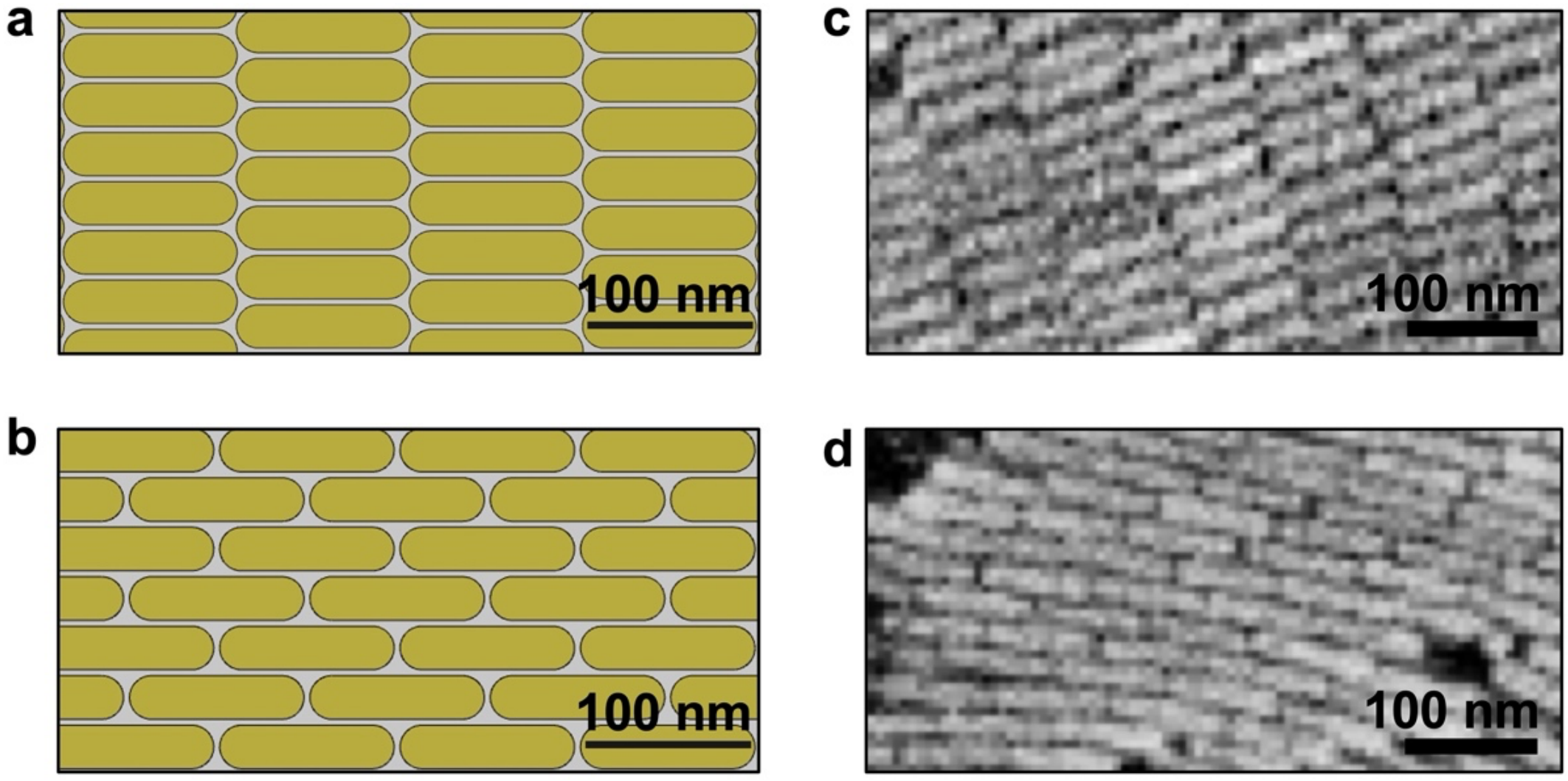

**Figure S15**. **Tuning thermal transport anisotropy through nanorod packing structure. a**, Simulation geometry for typical side-by-side close-packed centered rectangular packing structure, or "smectic-like" structure. **b**, "Brick wall pattern" with end-to-end packing with staggered adjacent rows. **c**, **d**, Selected SEM images of these two packing structures. We note that structure (**c**) is far more common in our nanorod SCs.

## SUPPLEMENTARY SECTION 10

### DIFFUSIVE HOPPING MODEL

We consider a two-dimensional diffusive hopping model of thermal transport between anisotropic nanocrystals to gain insight into the role of packing arrangement. This model assumes the limiting case in which the diffusive hopping holds for NC-to-NC transfer. We point out that this picture is not strictly correct because the true transfer involves a combination of rate-limited interfacial transfer and diffusion-limited intra-medium transport. Additionally, the three-dimensional arrangement of deeper supercrystal layers could modulate the anisotropy compared to the two-dimensional results below, but the general principles established here could be used to treat such specific cases.

Specifically, we take the limit that transport within the NC is instantaneous and transfer between the NCs is a Poisson process limited by the interfacial resistance with negligible transit time within the interstitial ligand medium. In this case, transport can be described by a diffusive random walk with site-to-site hop distances of $\Delta x$ and characteristic time $\Delta t$. In the continuum limit this leads to a diffusion coefficient along a given one-dimensional axis of $D = \Delta x^2/2\Delta t$. In an anisotropic system, hops between specific locations will have different distances $\Delta x_i$ and times $\Delta t_i$ such that the total diffusion coefficient along a given axis is the sum of the different contributions, $D = \sum_i (N_i/2)\Delta x_i^2/2\Delta t_i$ where $N_i$ is the number of acceptor NCs corresponding to $\Delta x_i$ and $\Delta t_i$ . We also note that all of the above considers the preasymptotic regime in which $D_{\|}/D_{\perp} \ll k_{NC}/dh_c$.

To find the thermal transport anisotropy in this limit, we then sum the contributions along each axis noting that some arrangements lead to additional contributions in the orthogonal directions (Figure S16). We simplify the problem by considering transfer between rectangles. For each hop, we take the distance to be the center-to-center distance, projected along each axis, and take the inter-NC spacing to be negligible. In terms of the interfacial thermal conductance $h_c$, the time constant in two-dimensions is given by $\Delta t = CA/h_c l$, where $C$ is the heat capacity, $A$ is the area common to all NCs (in place of the volume for a three-dimensional system) and $l$ is the linear surface length directly connecting two given nanocrystals (in place of the surface area through which heat transfers).

For the side-by-side centered rectangular packing in Figure S16c, transport along the major axis gives $N = 4$, $\Delta x = L$, and $l = d/2$. For the minor axis, transport to the adjacent NCs contributes $N = 2$, $\Delta x = d$ and $l = L$, but hops along the major axis also contribute to the minor axis direction with $N = 4$, $\Delta x = d/2$, and $l = d/2$. Taken together, we obtain the limiting thermal transport anisotropy for this packing geometry in two dimensions:

$$\frac{D_{\|}}{D_{\perp}} = \frac{L}{d}\frac{L/d}{L/d + 1/4}$$

This limit for the anisotropy is nearly equal to, but just below, the aspect ratio, approaching it for $L/d \gg 1/4$.

For the brick wall pattern in Figure S16d, transport along the major axis between end-to-end NCs gives $N = 2$, $\Delta x = L$, and $l = d$, and transfer across the minor axis also contributes to the major axis with $N = 4$, $\Delta x = L/2$ and $l = L/2$. Along the minor axis we have $N = 4$, $\Delta x = d$, and $l = L/2$. Taken together, we obtain the limiting thermal transport anisotropy for the brick wall pattern in two dimensions:

$$\frac{D_{\parallel}}{D_{\perp}} = \frac{L}{d}\left(1 + \frac{L}{4d}\right)$$

Here the anisotropy is always larger than the aspect ratio, consistent with simulations of this packing structure (Figure S16e) and the experimental findings for the bipyramids. To be precise, the anisotropy increases quadratically with aspect ratio (Figure S16e), motivating future explorations of this degree of freedom.

Note that for the limit $L/d = 1$ this model gives $D_{\parallel}/D_{\perp} = 4/5$ and $5/4$ due to hexagonal packing. This is an artifact of taking a geometry with six-fold symmetry and finding two orthogonal directions; the transport in that case actually has three axes of high diffusivity (along chains of NCs) and three axes of low diffusivity such that ultimate profile is effectively isotropic. The above model is relevant for $L/d > 1$ when two-fold symmetry sets in.

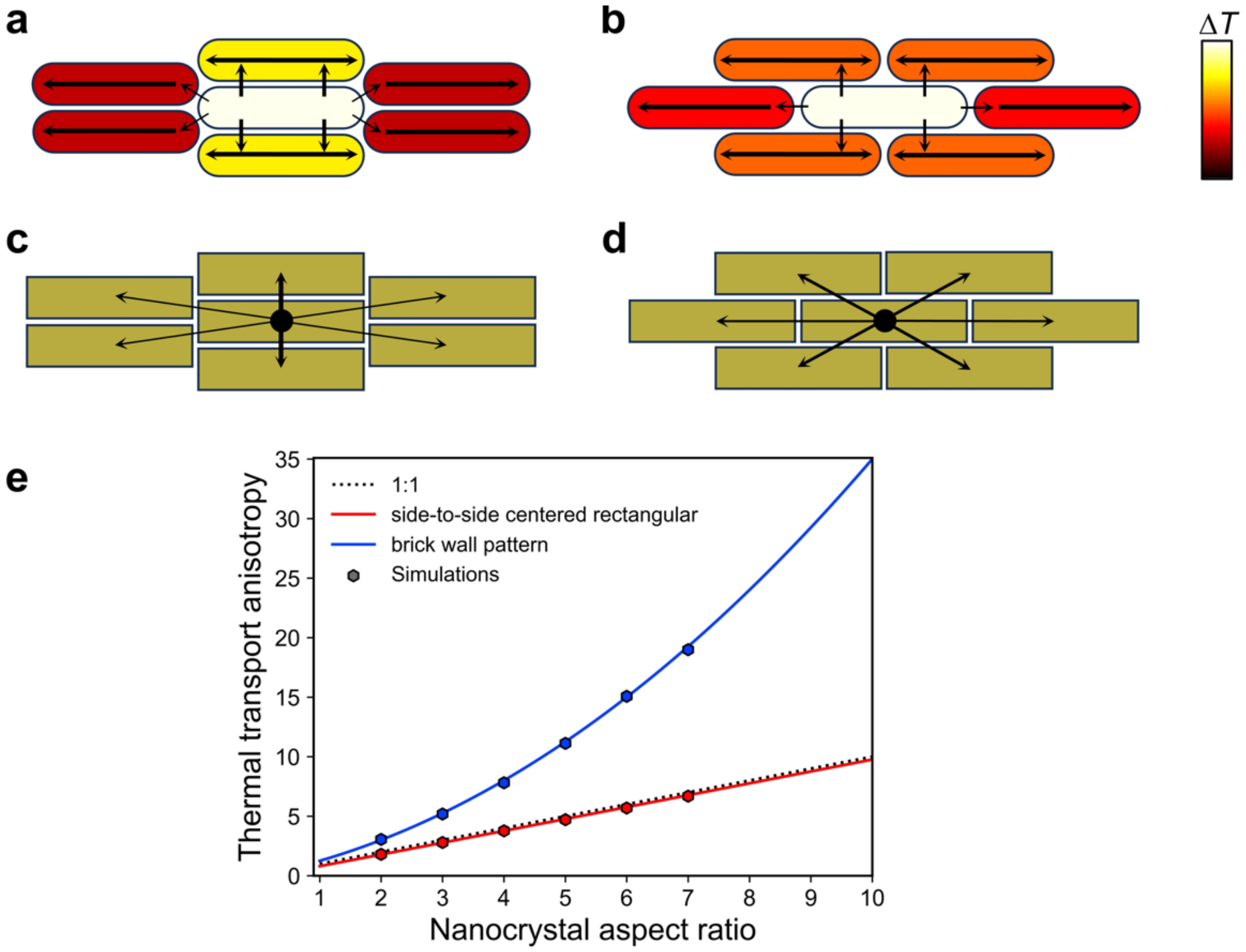

**Figure S16**. **Schematic depiction of diffusive hopping model enabling enhanced anisotropy for different two-dimensional packing geometries. a**,**b**, Cartoon of spatiotemporal picture in which, starting from a hot central NC, heat undergoes slow transfer across NC–NC boundary followed by rapid transport across the extent of the accepting nanocrystal. For side-by-side centered rectangular packing (**a**), transfer along the minor axis does not add to parallel transport, while for brick-wall pattern (**b**) transfer along the minor axis is followed by rapid expansion along the major axis over a distance of ~$L/2$. Relative temperatures are not quantitative and are for illustration purposes only. **c**,**d**, Cartoon of diffusive hopping picture in which heat undergoes hops of a distance $\Delta x$ with a characteristic time $\Delta t$ between the centers of the nanocrystals. In all panels the thickness of the arrow depicts the rate of the transfer. In panels (**c**) and (**d**) the length of the arrow corresponds to the distance of the hop. **e**, Plot of limiting thermal transport anisotropy from diffusive hopping model for the side-to-side centered rectangular packing arrangement in (**c**) and the brick wall pattern in (**d**). Finite element simulations are performed for arrangements (**c**) and (**d**) assuming no inter-NC spacing and an interfacial conductance of 40 MW $m^{-2}$ $K^{-1}$, matching with the hopping model in this limit.

## SUPPLEMENTARY SECTION 11

## INCREASED THERMAL TRANSPORT ANISOTROPY BY DECREASED SIDE-TO-SIDE NANOCRYSTAL COUPLING

### SEM image evidence of inter-nanocrystal voids

We explore the impact of inter-NC voids on reducing the thermal diffusivity along one axis of the SC assembly to enhance anisotropy. While controlling the side-to-side inter-NC spacing to include air gaps is experimentally challenging in nanorod SCs, we consider where inter-NC gaps might form naturally and perform measurements there. Examination of nanorod packing in SEM images here reveals instances in which the inter-NC spacing is larger than twice the ligand length (Figure below), implying the presence of inter-NC voids. While not discussed explicitly, examination of previous reports of nanorod SCs suggest such structure as well.[13,14] Such situations might occur, for example, when a row of adjacent nanorods redistribute around a missing nanorod or neighbors of polydisperse diameters, or due to the propagation of inter-nanorod spacing along the nanorod length when there are even small relative angles in smectic liquid-crystalline-type regions.

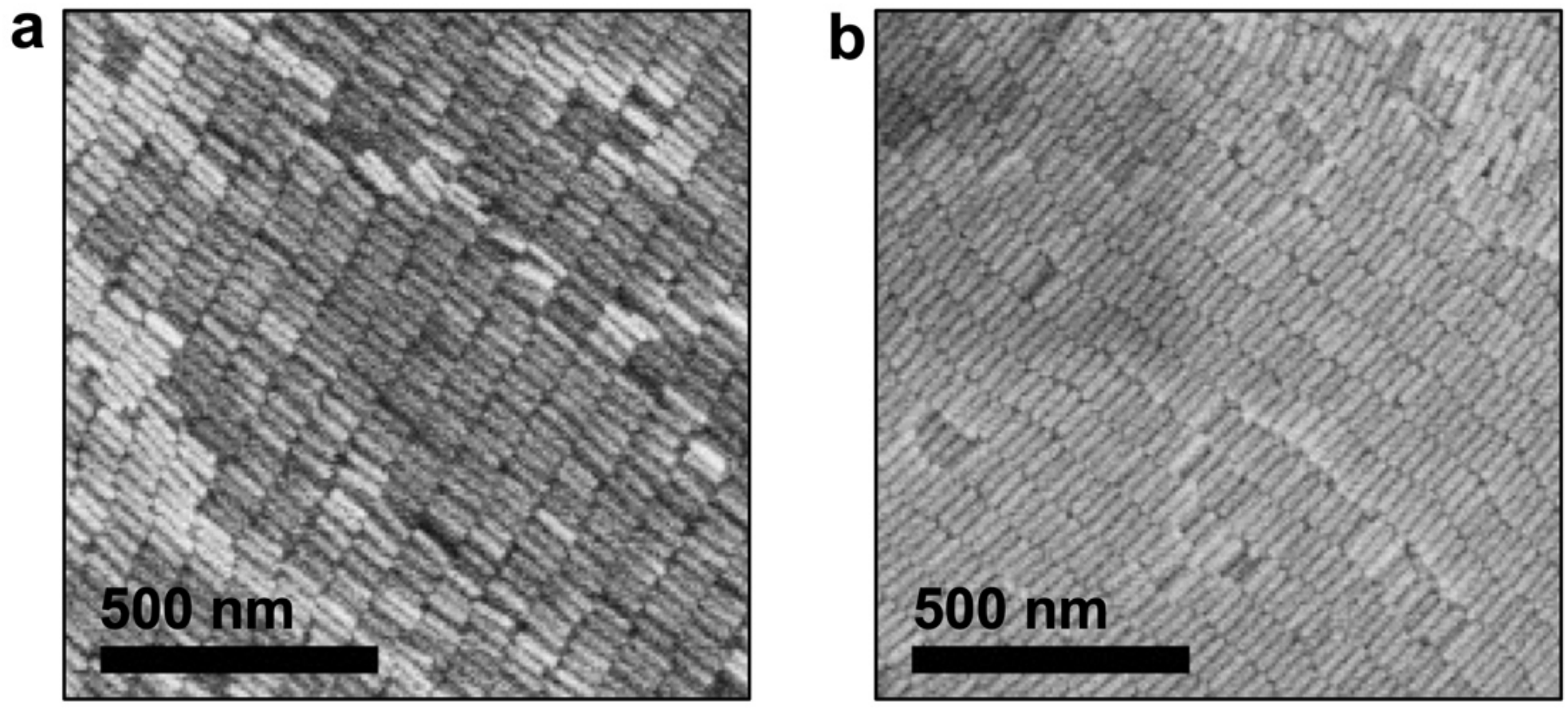

**Figure S17**. **Natural presence of minor axis voids.** Selected SEM images of (**a**) sparsely and (**b**) densely packed nanorod supercrystals. Several instances occur in (**a**) where the side-to-side inter-NC spacing is larger than twice the ligand length of 2.2 nm.

### Measurements on top and bottom layers of supercrystals

We performed spatiotemporal thermoreflectance measurements to find the anisotropic thermal diffusivity ratio in multiple locations of 4.0:1 and 5.6:1 nanorod SCs, probing from below (through the coverslip that acted as the substrate for self-assembly) and from above (by flipping samples over). Histograms of these thermal transport anisotropies and a table summarizing the underlying thermal diffusivities appear below. Under the assumption that bottom layers of SC are more likely to be well-packed with fewer inter-nanorod voids compared to the top layers of SC that may feature incomplete layers, data sets collected at the bottom SC layers at the SC–coverslip interface are designated as close-packed regions, and data sets collected at the top of SCs are designated as incomplete regions with voids. The optical penetration depth makes these reflection experiments moderately weighted by the surface behavior, leading to a measurable difference in the two cases.

**Table S6. Summary of effect on thermal diffusivity in regions with defects**

| Sample Title | $\langle D_{\parallel} \rangle$ (× $10^{-3}$ cm$^2$ s$^{-1}$) | $\langle D_{\perp} \rangle$ (× $10^{-3}$ cm$^2$ s$^{-1}$) | $\langle D_{\parallel}/D_{\perp} \rangle$ |
|---|---|---|---|
| 4.0:1 nanorods (bottom layers) | 5.0 ± 0.3 | 1.5 ± 0.2 | 3.2 ± 0.3 |
| 4.0:1 nanorods (top layers) | 2.5 ± 0.2 | 0.25 ± 0.05 | 10 ± 1 |
| 5.6:1 nanorods (bottom layers) | 2.7 ± 0.7 | 0.76 ± 0.13 | 4.1 ± 0.3 |
| 5.6:1 nanorods (top layers) | 2.7 ± 0.1 | 0.24 ± 0.02 | 11.2 ± 0.6 |

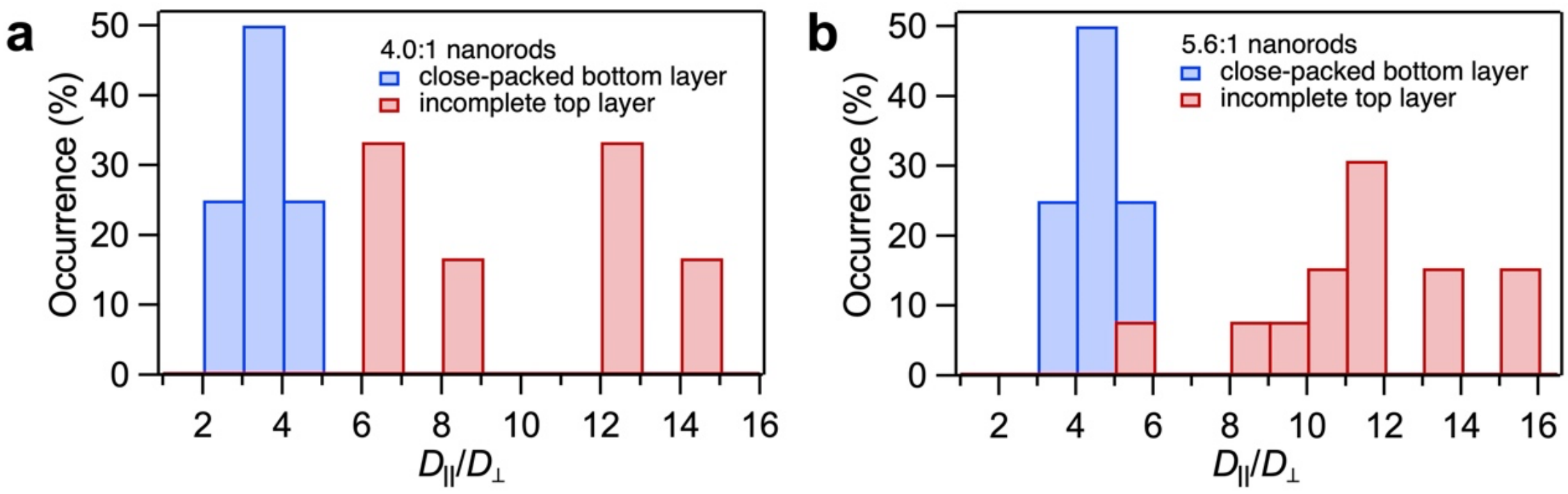

**Figure S18. Comparison of thermal transport anisotropy in close-packed versus incomplete regions of Au nanorod supercrystals.** Histograms for **a** 4.0:1 and **b** 5.6:1 Au–CTAB nanorod supercrystals of thermal diffusivity ratios $D_{\parallel}/D_{\perp}$ for multiple spatiotemporal thermoreflectance microscopy measurements collected with the beam reflecting from the bottom supercrystal-substrate interface (blue) and from the top (red).

**Finite element simulations of inter-nanocrystal voids**

We simulated a series of nanorod SCs with side-to-side voids. A ligand length of 2.2 nm was chosen as well as a closest inter-NC spacing of 3.4 nm in the major axis direction like the other simulations in this work, implying ligand overlap (interdigitation) of 1.0 nm. Starting from the side-by-side close-packed centered rectangular, "smectic-like" packing structure, the nanorods were displaced along the minor-axis, side-to-side direction to arbitrarily give voids of 2.4 nm while keeping a tip-to-tip spacing of 3.4 nm. This implies that the nanorods themselves interdigitate by slightly compressing the SC lattice constant along the major axis, consistent with the experimental result (Table S5) that the diffusivity along the minor axis is more strongly affected compared to the major axis. Simulations were performed as a function of nanorod aspect ratio holding the diameter fixed at 25 nm as a representative diameter of the nanorods studied here. In contrast to the case of no minor-axis voids (Figure 3), thermal diffusivity ratio scales superlinearly with nanorod aspect ratio, $D_{\parallel}/D_{\perp} > L/d$, as shown below. Similar to the brick wall pattern above but with a different origin, the hopping model predicts a quadratic dependence on aspect ratio. While the exact anisotropy is highly sensitive to the precise—and likely inhomogeneous—packing, this model scenario demonstrates the strong potential anisotropy achievable if side-to-side spacing were to be controlled.

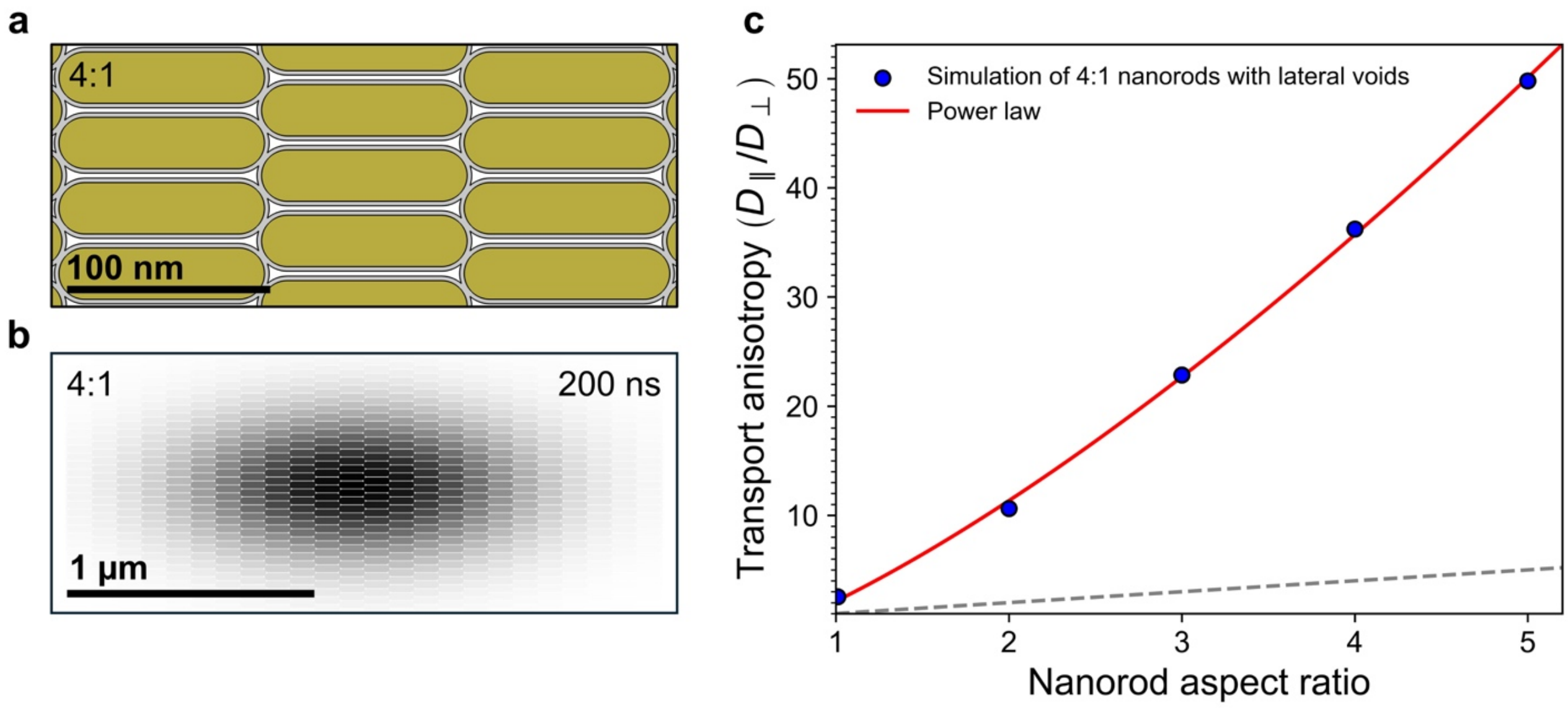

**Figure S19**. **Enhanced thermal anisotropy with side-to-side gaps. a**, Example simulation geometry of 4:1 nanorod supercrystal with voids introduced between the nanorod sides. **b**, Simulated temperature profile at 200 ns relative to the geometry in (**a**). **c**, Thermal transport anisotropy as a function of nanorod aspect ratio for a fixed nanorod diameter of 25 nm.

## SUPPLEMENTARY SECTION 12

### EFFECT OF LIGAND IDENTITY ON ANISOTROPY AND DIFFUSIVITY

In this work we have used Au NCs with a few different ligands including cetyltrimethylammonium bromide (CTAB), cetyltrimethylammonium chloride (CTAC), and polystyrene thiol (PSSH). These different ligands were chosen following previously established NC preparation and self-assembly protocols for each NC shape. To ensure that the different ligands do not affect our key findings when comparing different NC shapes with different ligands, we perform finite element simulations of the thermal diffusivity and thermal transport anisotropy for the different ligands.

We take 4:1 Au nanorods (25 nm × 100 nm) as a representative geometry and perform simulations using the different thermal parameters of the different ligands. Based on TEM images of the respective samples, we take inter-NC spacings of 4.4 nm for PSSH and 3.4 nm for CTAB/CTAC. While PSSH ligands are long and known to fill interstices, it is not clear for the CTAC/CTAB ligands and thus we compare both cases. For simulations we input the following parameters based on the literature: $\rho_{Au}$ = 19,300 kg m$^{-3}$, $C_{p,\,Au}$ = 129 J kg$^{-1}$ K$^{-1}$; $\rho_{CTAB/CTAC}$ = 500 kg m$^{-3}$,[3] $C_{p,\,CTAB/CTAC}$ = 1400 J kg$^{-1}$ K$^{-1}$;[4] $\rho_{PSSH}$ = 1000 kg m$^{-3}$,[5] and $C_{p,\,PSSH}$ = 1250 J kg$^{-1}$ K$^{-1}$ based on polystyrene thin films as a proxy.[6] The Au–CTAB/CTAC interface is treated as an equivalent thin resistive layer assuming an interfacial thermal conductance of 40 MW m$^{-2}$ K$^{-1}$ based on previous reports for Au nanocrystals with amine-based ligands.[15,16] The interfacial thermal conductance between the Au NCs and the PSSH ligands was estimated by assuming that the thermal transfer per ligand is governed by the Au–thiolate binding and ligand grafting density, $\sigma$. We take the known interfacial thermal conductance (220 MW m$^{-2}$ K$^{-1}$)[8] and ligand grafting density (4.8 ligands/nm$^2$)[8] for Au-dodecanethiol self-assembled monolayers (SAMs) and adjust for the reduced ligand coverage of the bulk polymer ligands—cited to be about 1 ligand/nm$^2$.[5] This assumption is based on previous work that showed that the interfacial thermal conductance $h_{NC\text{–}ligand}$ is proportional to ligand grafting density, which in turn is radius ($r$) or surface curvature dependent such that $h_{NC\text{–}ligand}(r)/h_{NC\text{–}ligand}(SAM) = \sigma(r)/\sigma(SAM)$.[8] Plugging in the values above gives $h_{NC\text{–}ligand} \approx 50$ MW m$^{-2}$ K$^{-1}$, which is the value used in our finite element simulations. Thermal transfer to the air (including voids) or substrate was neglected in the simulations.

The results of the simulations are given in Table S7. Comparing the geometries with filled interstices without voids, using the different ligand parameters leads to a small change in the absolute diffusivity and an inconsequential change in the thermal transport anisotropy. In fact, the possible presence of voids leads to a larger difference, but even in that case the change is quantitative but not qualitative. Overall, this suggests that, when comparing ligands with similar thermal and geometric properties as we do here, the choice of ligand has a small effect compared to the range of defects that are likely to be intrinsically present in NC SCs.

**Table S7. Simulations of diffusivity and anisotropy for different ligands**

| Ligand | $\langle D_{\parallel} \rangle$ (× 10$^{-3}$ cm$^2$ s$^{-1}$) | $\langle D_{\perp} \rangle$ (× 10$^{-3}$ cm$^2$ s$^{-1}$) | $\langle D_{\parallel}/D_{\perp} \rangle$ |
|---|---|---|---|
| CTAB/CTAC with voids | 4.9 | 1.7 | 3.0 |
| CTAB/CTAC without voids | 6.1 | 1.7 | 3.6 |
| PSSH without voids | 5.5 | 1.6 | 3.5 |